\documentclass[twocolumn]{aastex631}
\pdfoutput=1 
\usepackage[caption=false]{subfig}  
\usepackage{comment}
\usepackage{tabularx}

\newcommand\N[1]{{\color{blue} \bf #1}} 
\usepackage{soul}

\shorttitle{Rebrightening of extragalactic transients}
\shortauthors{Soraisam et al.}
\graphicspath{{./}{figures/}}

\newcolumntype{b}{X}
\newcolumntype{S}{>{\hsize=.4\hsize}X}
\newcolumntype{s}{>{\hsize=.3\hsize}X}

\begin{document}

\title{Optical Rebrightening of Extragalactic Transients from the Zwicky Transient Facility}

\correspondingauthor{Monika Soraisam}
\email{monika.soraisam@noirlab.edu}

\author[0000-0001-6360-992X]{Monika~Soraisam}
\affiliation{NSF's National Optical-Infrared Astronomy Research Laboratory, Gemini North, 670 N Aohoku Place, Hilo, HI 96720, USA}
\author[0000-0001-6685-0479]{Thomas~Matheson}
\author[0000-0003-1700-5740]{Chien-Hsiu~Lee}
\author{Abhijit~Saha}
\affiliation{NSF's National Optical-Infrared Astronomy Research Laboratory, 950 N Cherry Ave, Tucson, AZ 85719, USA}
\author{Gautham~Narayan}
\affiliation{Department of Astronomy, University of Illinois at Urbana-Champaign, Urbana, IL 61801, USA}
\author{Nicholas~Wolf}
\author{Adam~Scott}
\affiliation{NSF's National Optical-Infrared Astronomy Research Laboratory, 950 N Cherry Ave, Tucson, AZ 85719, USA}
\author{Stephanie~Figuereo}
\author{Rafael~Nu\~{n}ez}
\author[0000-0001-7494-5910]{Kevin~McKinnon}
\author[0000-0001-8867-4234]{Puragra~Guhathakurta}
\affiliation{Department of Astronomy \& Astrophysics, University of California, Santa Cruz, 1156 High St, Santa Cruz, CA 95064, USA}
\author{Thomas~G.~Brink}
\affiliation{Department of Astronomy, University of California, 501 Campbell Hall, Berkeley, CA 94720-3411, USA}
\author[0000-0003-3460-0103]{Alexei~V.~Filippenko}
\affiliation{Department of Astronomy, University of California, 501 Campbell Hall, Berkeley, CA 94720-3411, USA}
\affiliation{Miller Institute for Basic Research in Science, University of California, Berkeley, CA 94720, USA}
\author[0000-0001-5510-2424]{Nathan~Smith}
\affiliation{Steward Observatory, University of Arizona, 933 N.~Cherry Ave., Tucson, AZ 85721, USA}

\begin{abstract}
Ongoing large-scale optical time-domain surveys, such as the Zwicky Transient Facility (ZTF), are producing alerts at unprecedented rates. Analysis of transient sources has so far followed two distinct paths: archival analysis of data on transient sources at a time when they are no longer observable and real-time analysis at the time when the sources are first detected. The latter is the realm of alert brokers such as the Arizona-NOIRLab Temporal Analysis and Response to Events System (ANTARES). In this paper, we synthesize the two analysis paths and present a first systematic study of archival alert-broker data, focusing on extragalactic transients with multipeaked light curves identified in the ANTARES archive of ZTF alerts. Our analysis yields a sample of 37 such sources, including core-collapse supernovae (with two analogs of iPTF14hls), thermonuclear supernovae interacting with their surrounding circumstellar medium, tidal disruption events, luminous blue variables, and as yet unclassified objects. A large fraction of the identified sources is currently active, warranting allocation of follow-up resources in the immediate future to further constrain their nature and the physical processes at work.
\end{abstract}


\section{Introduction} \label{sec:intro}
Wide-area large-scale surveys in the time domain (e.g., ASAS-SN, \citealt{Shappee-2014}; ATLAS, \citealt{Tonry-2018}; Zwicky Transient Facility [ZTF], \citealt{Bellm-2019}) are increasingly pushing the frontier of astrophysical discoveries. There has been a significant transformation in the extant methodology of performing real-time science in time-domain astronomy motivated by the data deluge of the ongoing ZTF survey. To this end, software instruments, in particular alert-brokers \citep[e.g.,][]{Saha-2016, Forster-2021, Matheson-2021}, that filter the enormous data to a manageable subset of the most interesting astrophysical events, are becoming an integral component. For the upcoming Vera Rubin Observatory Legacy Survey of Space and Time (LSST, \citealt{LSST}), such specialized infrastructure will in fact be indispensable.            

Early-time observations of transient events have proven to be effective in providing insights into various astrophysical processes, such as shock-breakout in Type II supernovae (SNe; \citealt{Bersten-2018}), possible companion interaction in SNe~Ia \citep{Dimitriadis-2019}, etc. Equally powerful are the late-time observations of these explosive phenomena. Such observations have revealed peculiar events like nonterminal or impostor SNe that may precede actual SNe  \citep[SN 2009ip;][]{Smith-2010,Mauerhan-2013} and enduring multipeaked light curves as in the unusual SN iPTF14hls \citep{Arcavi-2017}. These rare discoveries, centered on the evolution of SNe over long timescales, have been pivotal in establishing our lack of understanding of key physical processes operating in massive-star progenitors. For example, mass loss in these stars has been plagued by uncertainties --- its driving mechanisms in the early or late stages of nuclear burning are to-date not well understood \citep{Smith-2014}. However, often mass loss in these stars is the suspect for anomalous features seen in their SN explosions \citep{Smith-2017}. Precursor outbursts seen in some SNe (e.g., SN 2010mc, \citealt{Ofek-2013}; SN 2009ip, \citealt{Smith-2010}) have been attributed to eruptive mass-loss events in the progenitors, while interaction of SN ejecta and circumstellar material (CSM) built up by episodes of mass loss has been invoked to explain dramatic late-time bumps seen in the light curve of SN iPTF14hls \citep{Andrews-2018}.

It is clear that uncovering similar events will contribute to advancing our knowledge of these extreme phenomena and stellar evolution in general. Here, we present our results from mining the ZTF survey data with the main objective to find extragalactic transients (comprising largely SNe) with {\it bumpy} light curves --- i.e., those with multiple peaks. In particular, we use the archival alert database of the \texttt{ANTARES} event broker \citep{Matheson-2021} in conjunction with the NOIRLab science platform Astro Data Lab{\footnote{https://datalab.noirlab.edu/}}. Many of the events we have discovered are active at the time of writing; thus, they can be monitored by the community to constrain their physical processes.   

The paper is organized as follows. We describe our sample selection and modeling of the light curves in Sects.~\ref{sec:sample} and \ref{sec:lightcuve}, respectively. Our results are shown and discussed in Sect.~\ref{sec:results}, and we conclude in Sect.~\ref{sec:conclude}.

\section{Selection of events} \label{sec:sample}
In order to find bumpy 
extragalactic events, we use the output of one of the filters running on \texttt{ANTARES}, viz. {\it HighAmp}{\footnote{The code is public and can be accessed at \url{https://antares.noirlab.edu/pipeline}}}. This is a simple filter to flag alerts from ZTF in real time displaying a significant amplitude. The amplitude is measured as the error-weighted root-mean-square (RMS) deviation about the mean of the light curve, and the threshold for flagging is set at 0.5~mag in any passband ($g$ and $r$ for ZTF). The output of this filter bifurcates into a transient and a variable star stream based on the absence or presence of a nearby (within $1.5''$) point-like source in the template image used for subtraction. The transient stream comprises mostly extragalactic events such as SNe and active galactic nuclei (AGNs), and to a lesser extent of Galactic events (mainly dwarf novae and classical novae). We therefore use this stream for our purpose. 

There were approximately 14,500 unique sources (or loci) in the {\it high\_amplitude\_transient\_candidate} output stream of \texttt{ANTARES} through 2021 October 15, which was the freeze date for this analysis. We consider this as our {\it master sample}. We post-process the loci in this sample on the Astro Data Lab science platform, which provides a JupyterLab environment and access to the archival \texttt{ANTARES} alert database including all the value-added products, such as annotations from cross-matching with various legacy catalogs (see \citealt{Matheson-2021} for details). In particular, we apply the following three selection criteria: (i) at least 10 detections in any of the ZTF passbands, (ii) all detections are positive, i.e., brightening events as compared to the template image used in subtraction, and (iii) no counterparts in the Million Quasars catalog \citep{Milliquas-2021} to filter out variability of AGNs. This resulted in 2089 loci, $\sim 14$\% of the master sample size, which forms our penultimate sample.

We smooth the light curves of this penultimate sample and select events with multiple peaks, as described in the next section. The remaining light curves are then visually vetted to identify extragalactic explosive transients with bumpy light curves.

\section{Light curve modeling} \label{sec:lightcuve}
We use Gaussian Process (GP) regression to model the light curves of the loci from Sect.~\ref{sec:sample}. GP allows for a continuous interpolation of the light curves by modeling their correlation function in a parametric way. 
To this end, we use \texttt{celerite} \citep{celerite1, celerite2}, which provides a scalable GP implementation, ideal for the thousands of light curves in our sample. 

\begin{figure*}
    \centering
    \includegraphics[width=80mm]{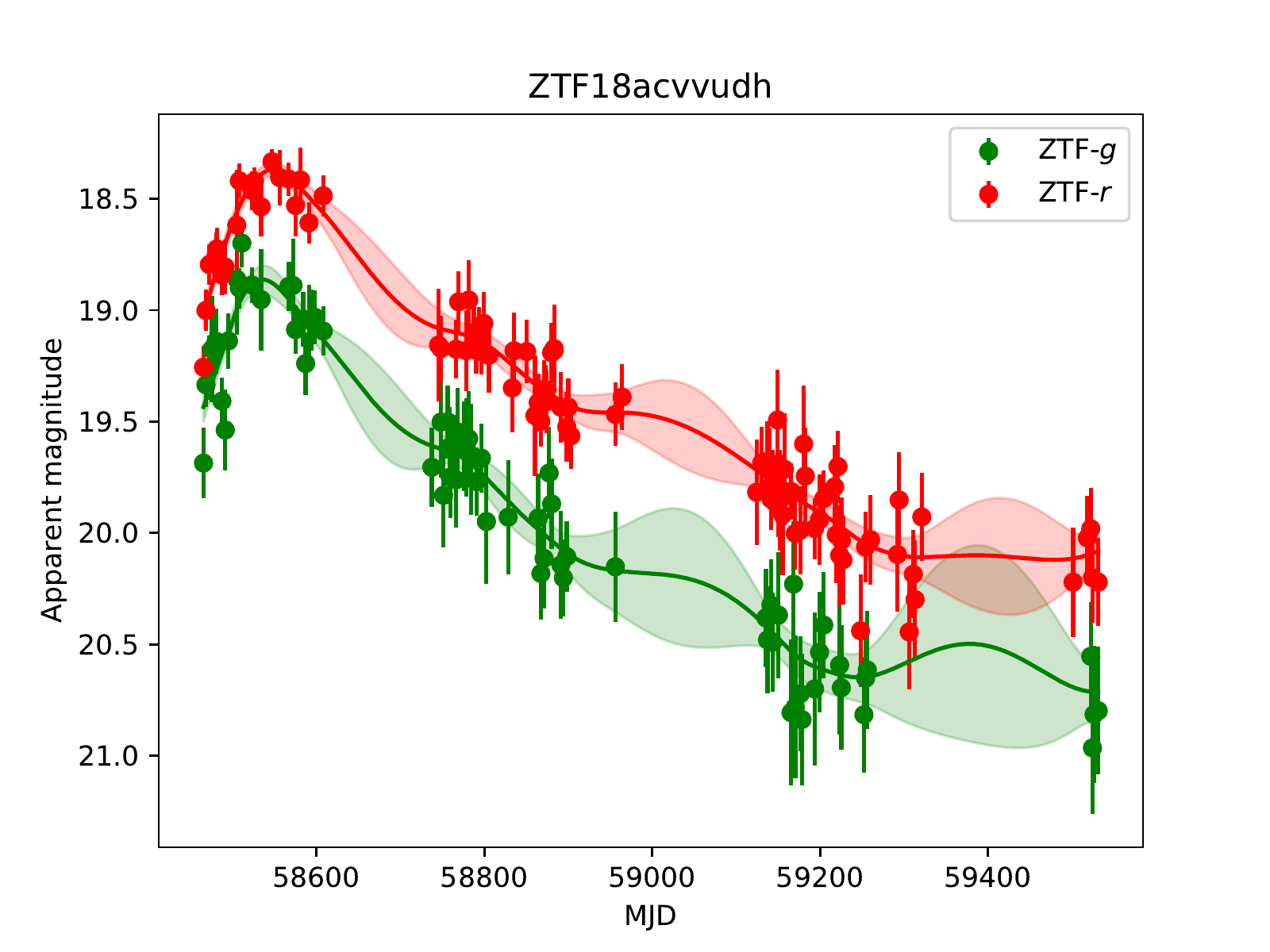}\hfill\includegraphics[width=80mm]{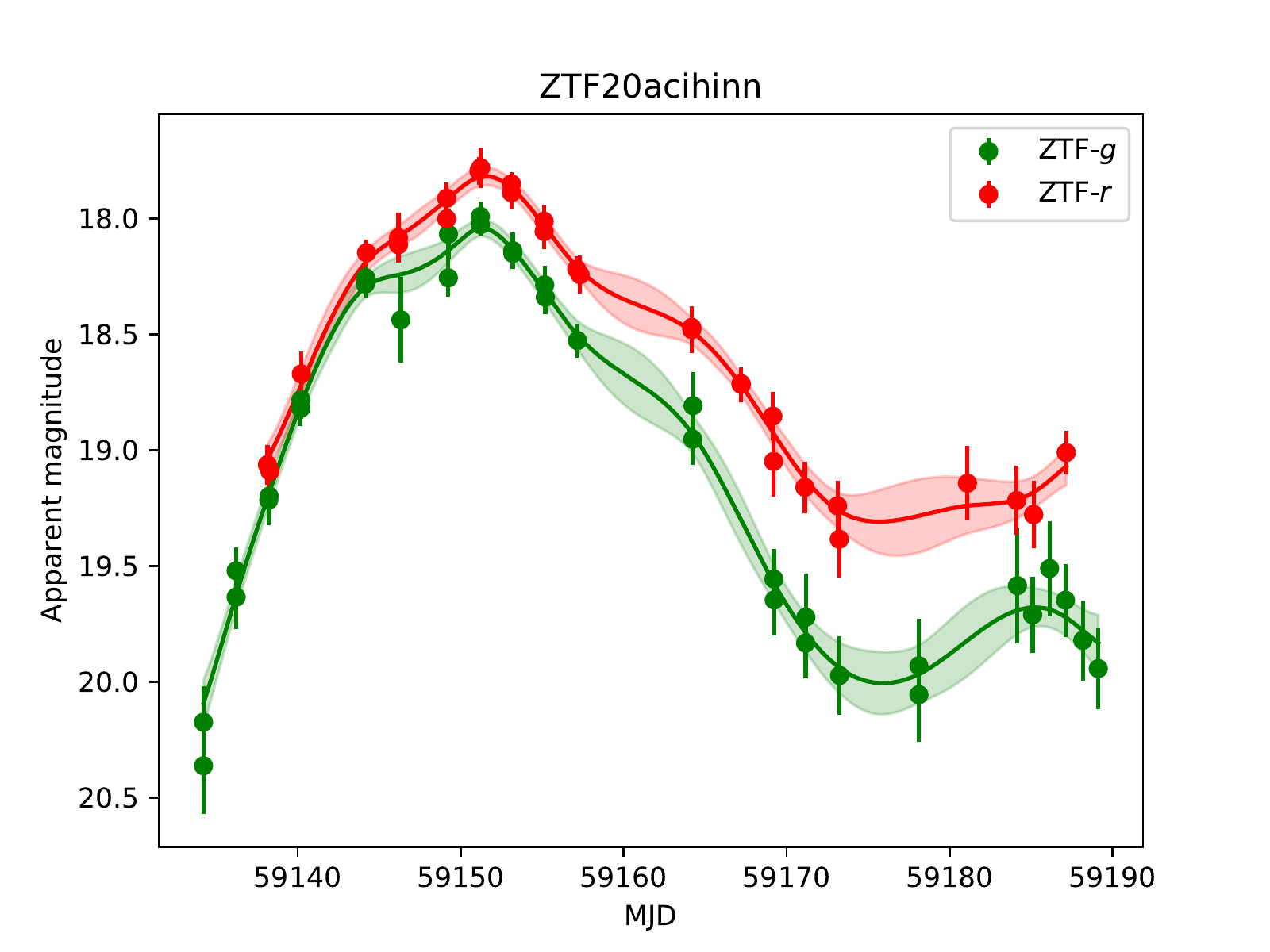}
    \includegraphics[width=80mm]{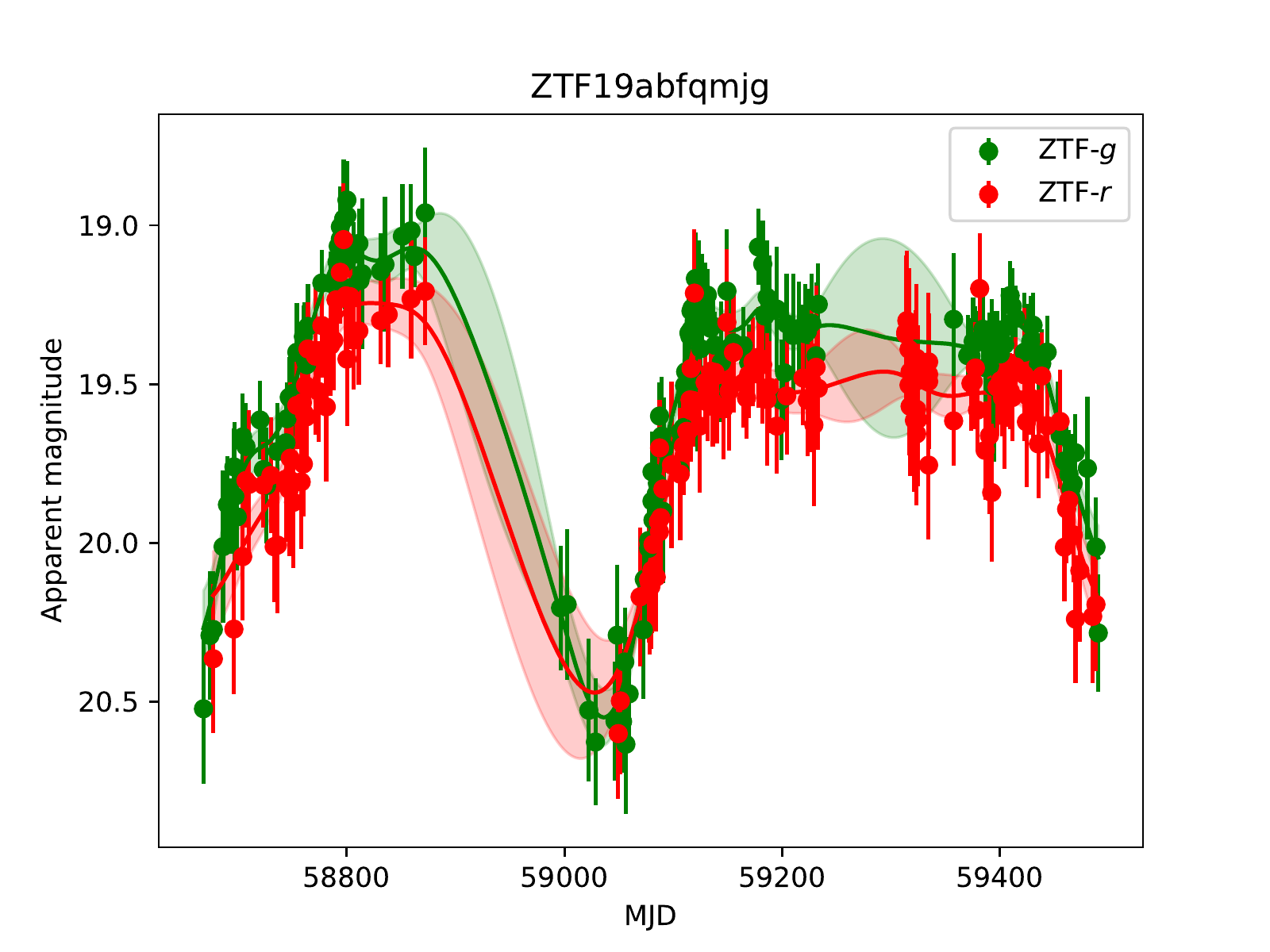}\hfill\includegraphics[width=80mm]{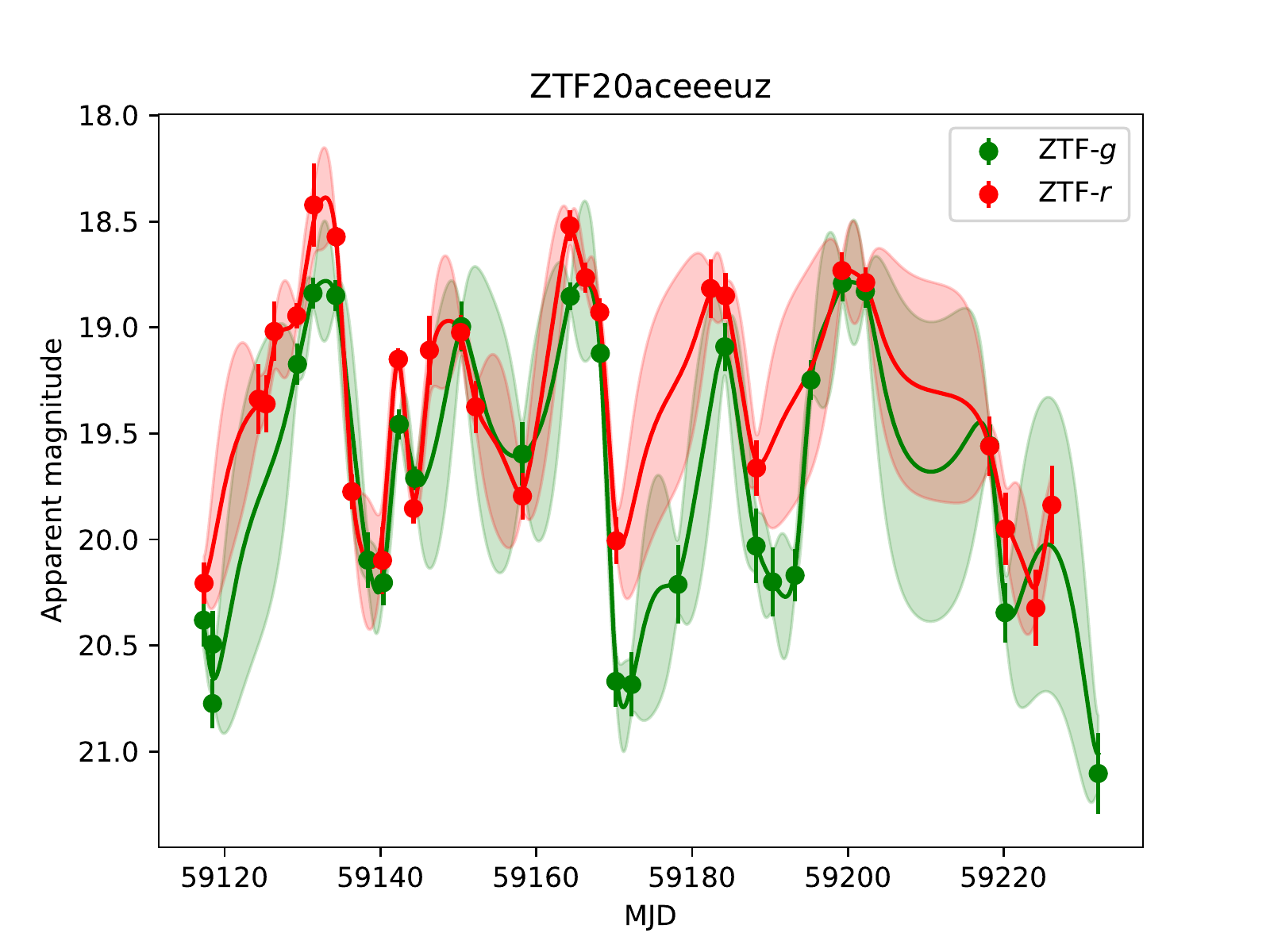}
    \includegraphics[width=80mm]{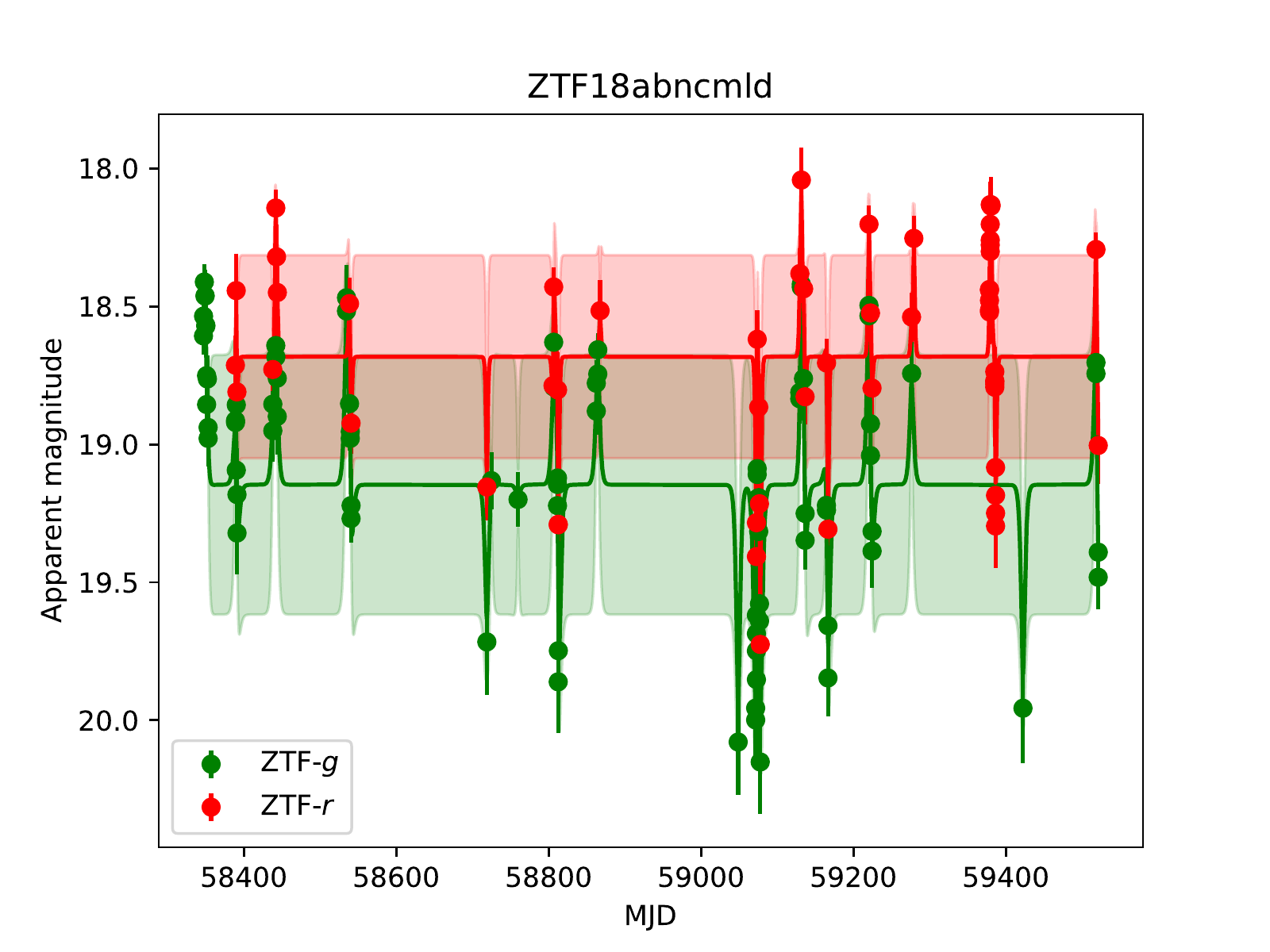}
    \caption{Example smoothed ZTF light curves. The five loci were chosen to highlight the quality of the \texttt{celerite} GP regression, which is represented by the solid lines. The green and red shaded regions show the $1\sigma$ uncertainty of the regression for the ZTF $g$ and $r$ passbands, respectively. 
    The top two panels (ZTF18acvvudh, ZTF20acihinn) are SN candidates. In the middle panels, ZTF19abfqmjg is a nuclear transient candidate, while ZTF20aceeeuz has been classified as a nova in M33 \citep{Reguitti-2020}. The bottom panel is a Galactic dwarf nova \citep{Watson-2006}. As is evident from the plots, the GP regression performs well for the extragalactic events, but fails for the dwarf nova.}
    \label{fig:exlcs}
\end{figure*}

From the available covariance functions in the \texttt{celerite} library, we opted for the series approximation of a Matern-3/2 function, parameterized by an amplitude and a timescale. We verified visually that this kernel choice returns a reasonable smoothing for most extragalactic events, although it seems to fail for some, which are mainly dwarf-nova contaminants (see Fig.~\ref{fig:exlcs}). These edge cases can typically be identified by large uncertainty values ($\sigma$) returned by \texttt{celerite}, which also tend to be similar from time point to time point. We use this information to remove the affected loci by thresholding on the uncertainty values, dropping events for which the
median $\sigma$-value is greater than 80\% of the maximum $\sigma$-value.

Finally, we apply the peak-finding algorithm of \texttt{scipy}. We require the light curves in any of the two ZTF passbands to have at least one peak with a {\it prominence} (height between the peak and its surrounding baseline{\footnote{\url{https://docs.scipy.org/doc/scipy/reference/generated/scipy.signal.peak_prominences.html}}}) of greater than 0.5~mag over a timescale or width greater than 10~days. This is to ensure the presence of a prominent (i.e., SN-like) outburst in the light curve. It must {\it additionally} have at least one peak with a prominence greater than 0.1~mag over a width greater than 5~days. This gives us 223 loci with multiple peaks. To remove contamination from variable stars, which may be too faint in archival images used in discerning the two $HighAmp$ stream (see Sect.~\ref{sec:sample}), we filter out events with matches either in the Guide Star Catalog \citep{Lasker-2008}, the ASAS-SN catalog of variable stars \citep{Jayasinghe-2019}, or the Catalina Surveys variable star catalog \citep{Drake-2014} using \texttt{ANTARES} annotation. Our final sample comprises 205 events, which is just $1.4\%$ of the master sample size. 78 of these have been typed as either SN, tidal disruption event (TDE), AGN, or luminous blue variable (LBV) on the Transient Name Server{\footnote{\url{https://www.wis-tns.org/}}}. 

It is to be noted that we have chosen rather permissive selection criteria above in favor of completeness, particularly for finding multiple peaks, since the photometric uncertainty can be up to 0.1~mag at the faint end, near the limit of detection of ZTF ($\sim 20\mbox{--}21$~mag; \citealt{Masci-2019PASP}). We therefore further extract multiple features using the Python package feATURE eXTRACTOR FOR tIME sERIES (\texttt{feets}; \citealt{feets-2018}) for these 205 events to investigate contaminants using the distributions of these features. We do not, however, find any significant trend that would allow us to further reduce the size of our sample. Hence, we resort to visual vetting for selecting the sources with the most clearly identifiable secondary peaks, as described in the next section.

\section{Results} \label{sec:results}
\begin{figure}
    \centering
    \includegraphics[width=80mm]{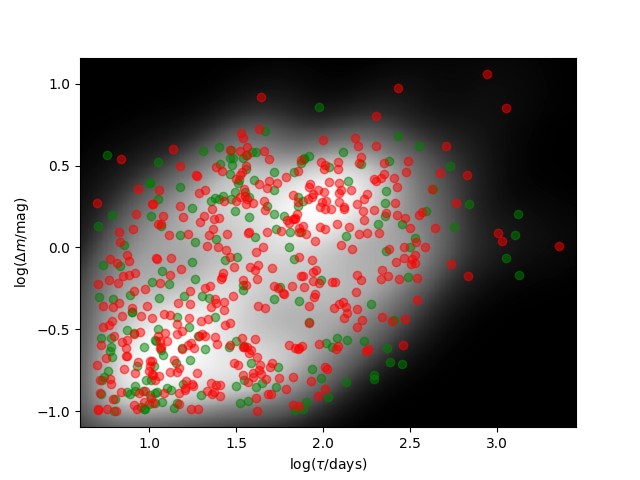}
    \caption{Variation of the prominence of the peaks, $\Delta m$, against their corresponding widths, $\tau$, for the final sample of loci. The green and red points indicate the results for the ZTF $g$ and $r$ passbands, respectively, while the background shading shows the density of these points (brighter regions have higher density values).}
    \label{fig:width_prom}
\end{figure}

The variation of the strength or prominence of the peak against its width for our final sample of events is shown in Fig.~\ref{fig:width_prom}. A general trend of increasing peak strength with peak width is evident from the plot. This is likely indicative of the fact that our final sample is dominated by extragalactic transients, mainly SNe, whose light curves are known to show a similar trend.

We visually examine the light curves of the 205 sources in the final sample obtained in Sect.~\ref{sec:lightcuve} and identify 37 with prominent bumps in their light curves, mostly corresponding to episodes of rebrightening at late times. These sources are listed in Table~\ref{tab:result}. We display the light curves of nine of these sources spanning the various categories of extragalactic transients (see below) in Fig.~\ref{fig:bumps}; the rest of the sources are shown in Appendix~\ref{appendix:bumpies}. Some of them are currently active (e.g., SN 2019meh, SN 2020xkx, LBV 2021blu, TDE 2020acka), and can therefore be further observed to help unravel their late-time physics.

\begin{figure*}
    \centering
    \includegraphics[width=80mm]{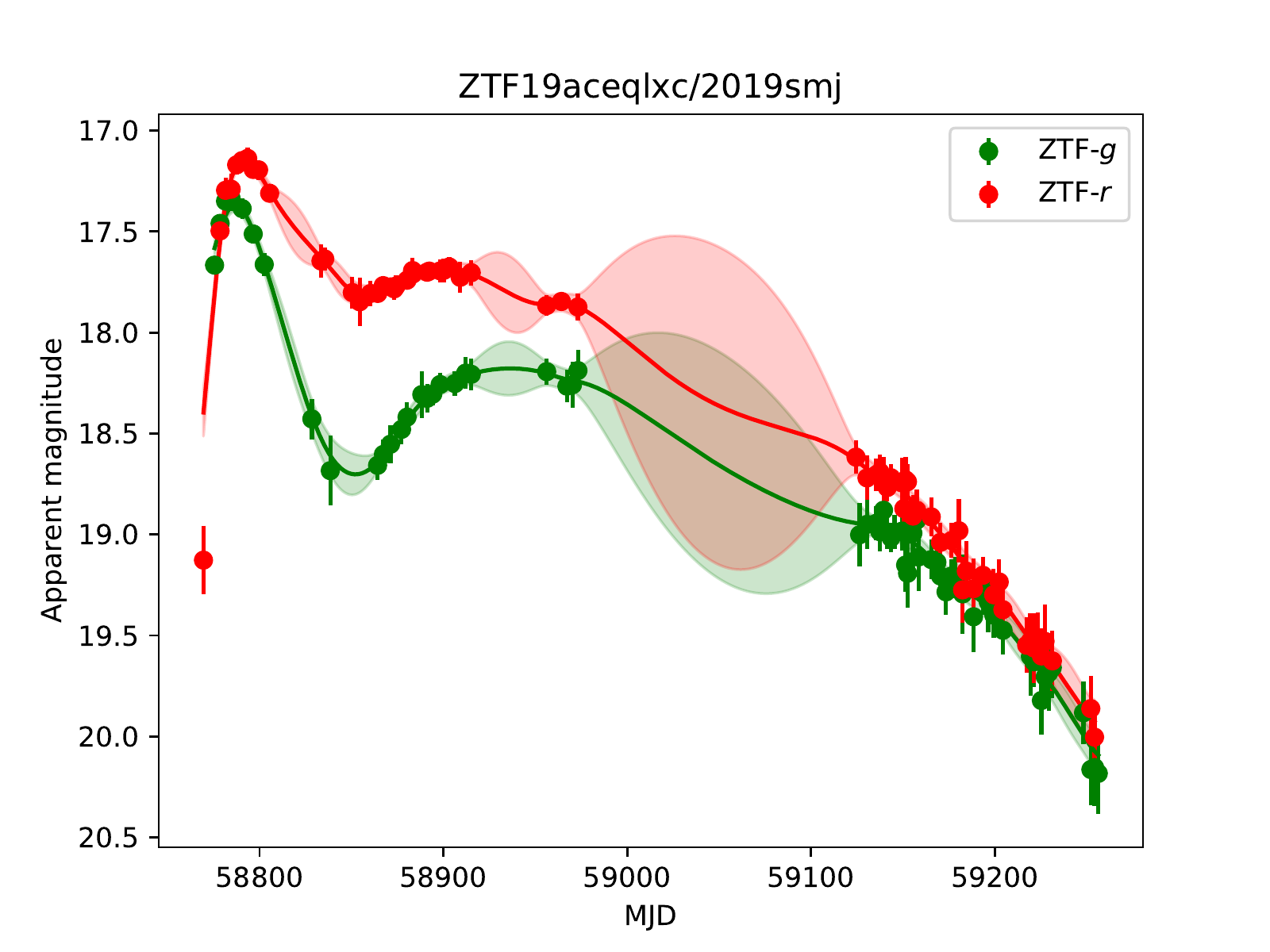}\hfill\includegraphics[width=80mm]{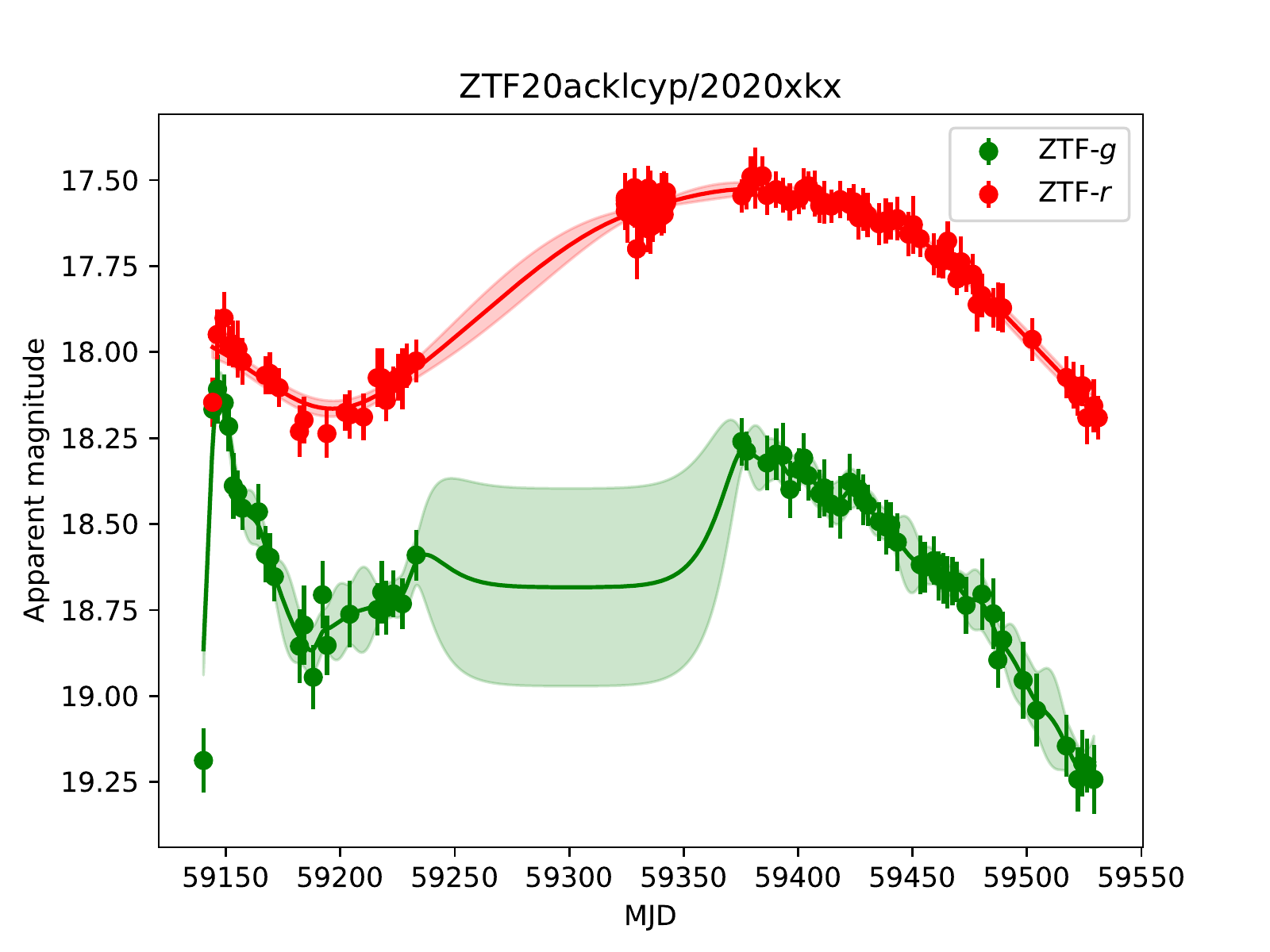}
    \includegraphics[width=80mm]{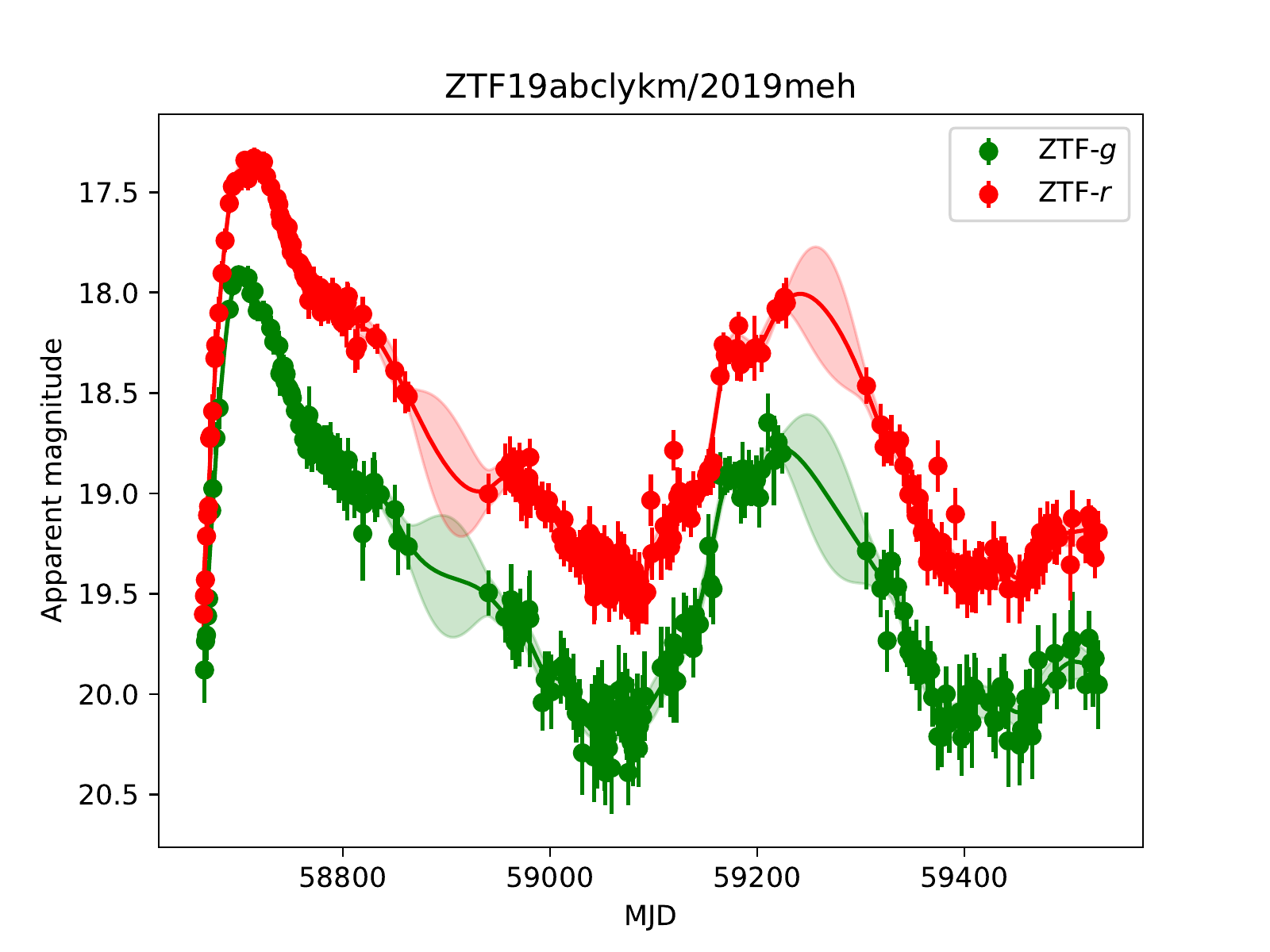}\hfill\includegraphics[width=80mm]{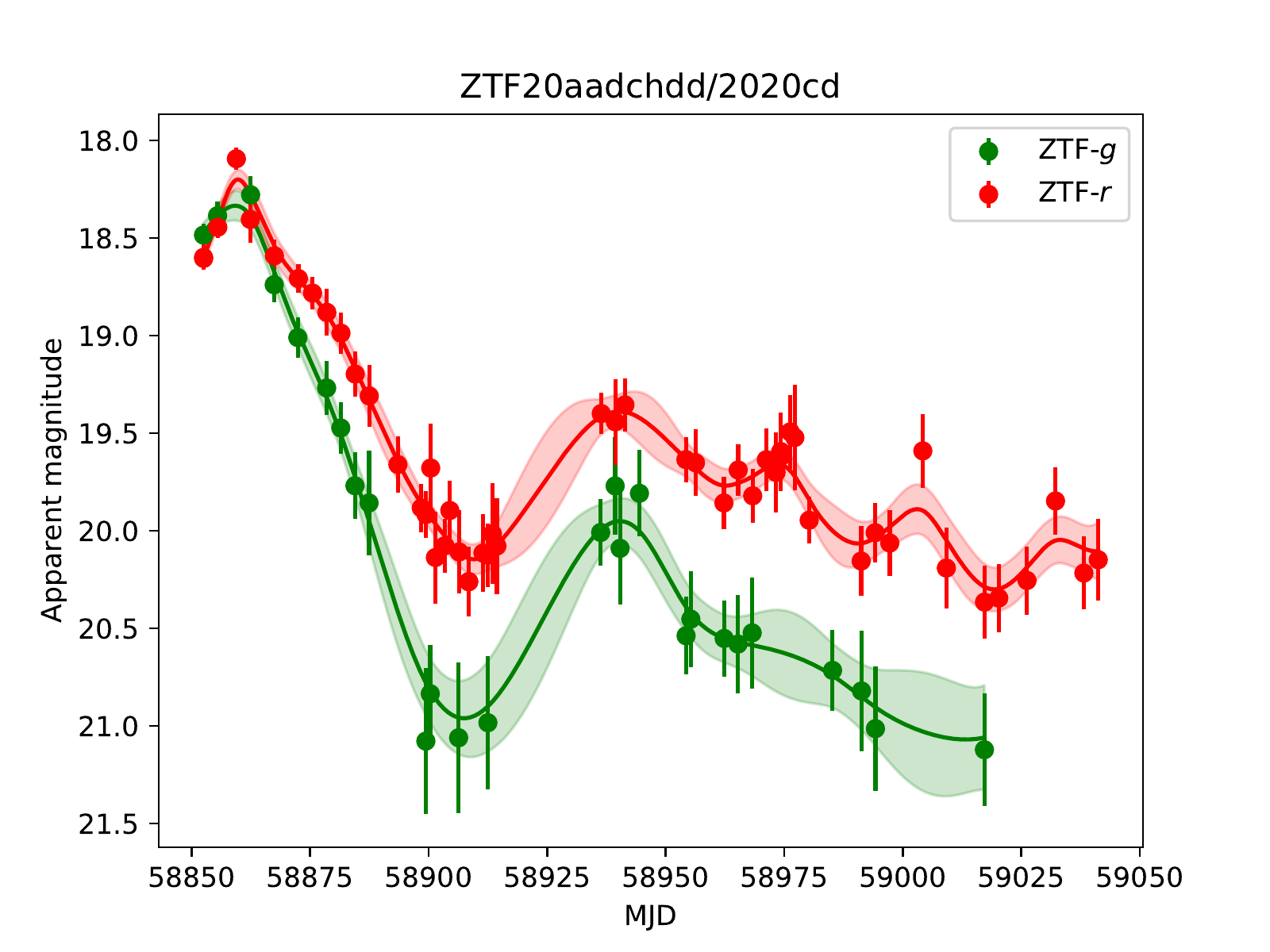}
    \includegraphics[width=80mm]{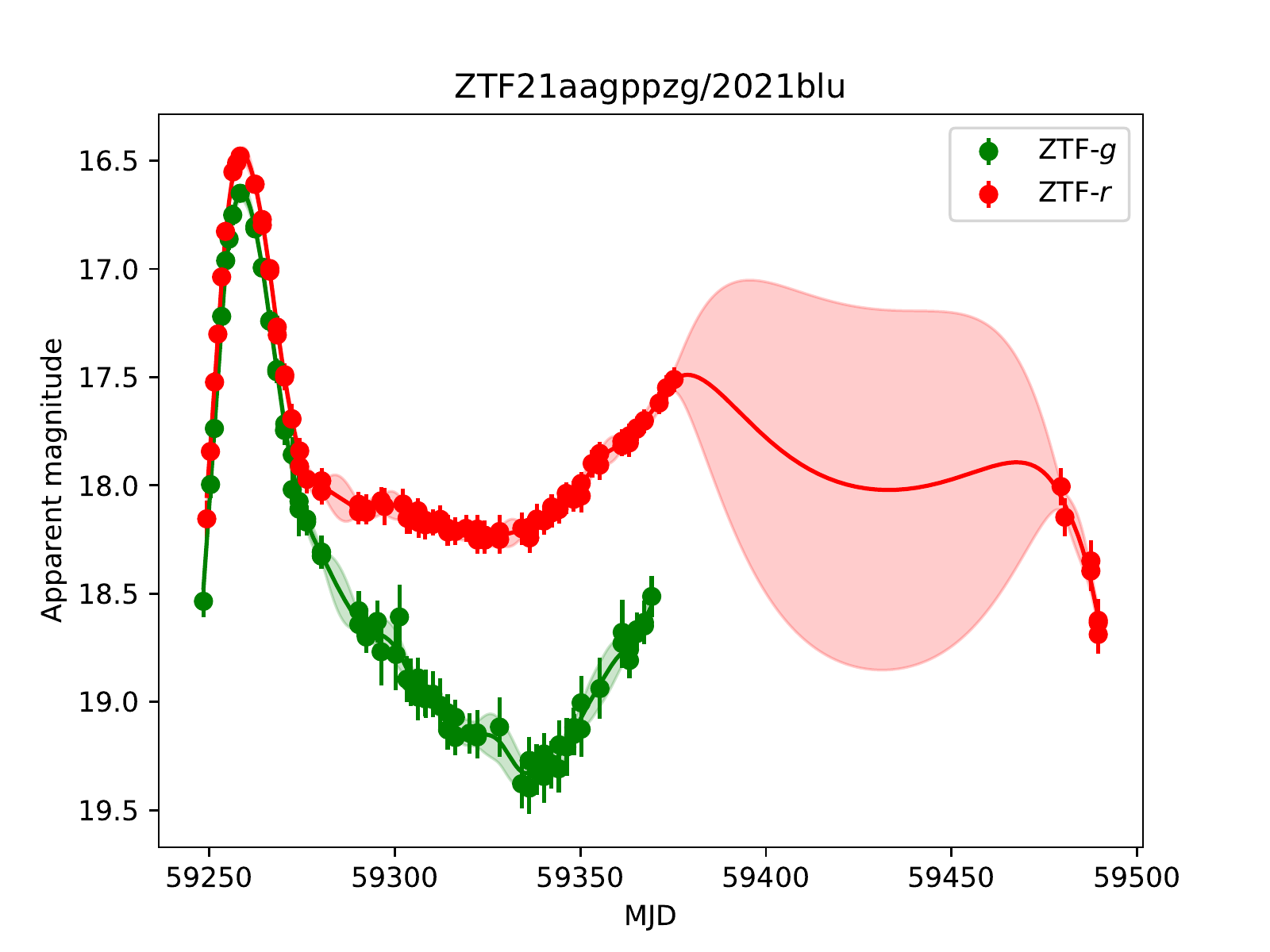}\hfill\includegraphics[width=80mm]{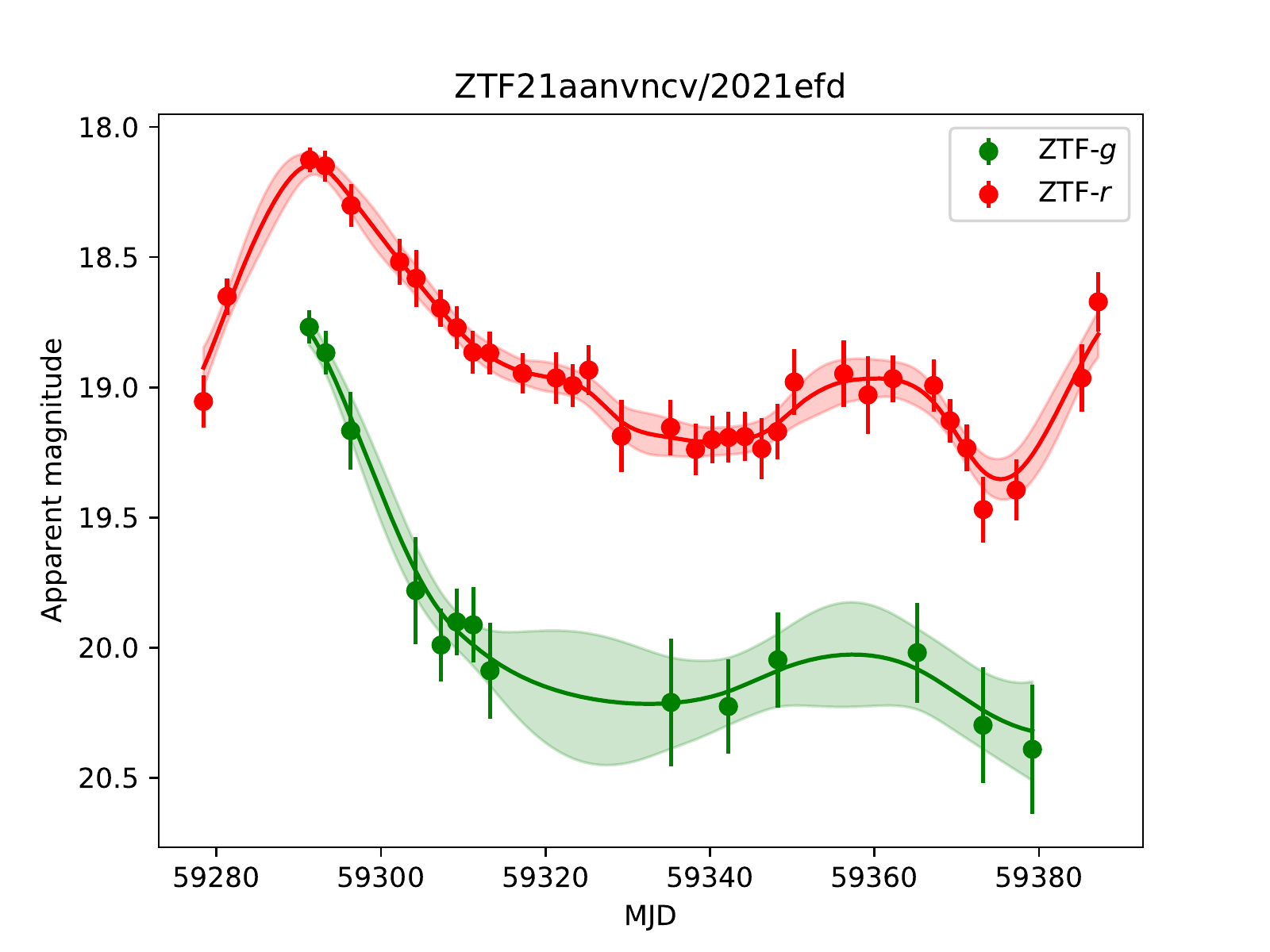}
    \caption{ZTF light curves of bumpy extragalactic transients. The lines and symbols are the same as in Fig.~\ref{fig:exlcs}. 
    The top two panels (ZTF19aceqlxc/2019smj, ZTF20acklcyp/2020xkx) are SNe of Type IIn \citep{Fremling2019TNSCR2297,Gromadzki2020TNSCR3243}. In the middle panels, ZTF19abclykm/2019meh is a Type II SLSN \citep{Nicholl2019TNSCR1586}, and ZTF20aadchdd/2020cd an SN~II \citep{Dahiwale2020TNSCR.204}. In the bottom panels, ZTF21aagppzg/2021blu is an LBV in the Spider galaxy \citep{Uno2021TNSCR.393}, and ZTF21aanvncv/2021efd an SN~Ib/c \citep{Reguitti2021TNSCR.682}.}
    \label{fig:bumps}
\end{figure*}

\begin{figure*}
    \ContinuedFloat
    \centering
    \includegraphics[width=80mm]{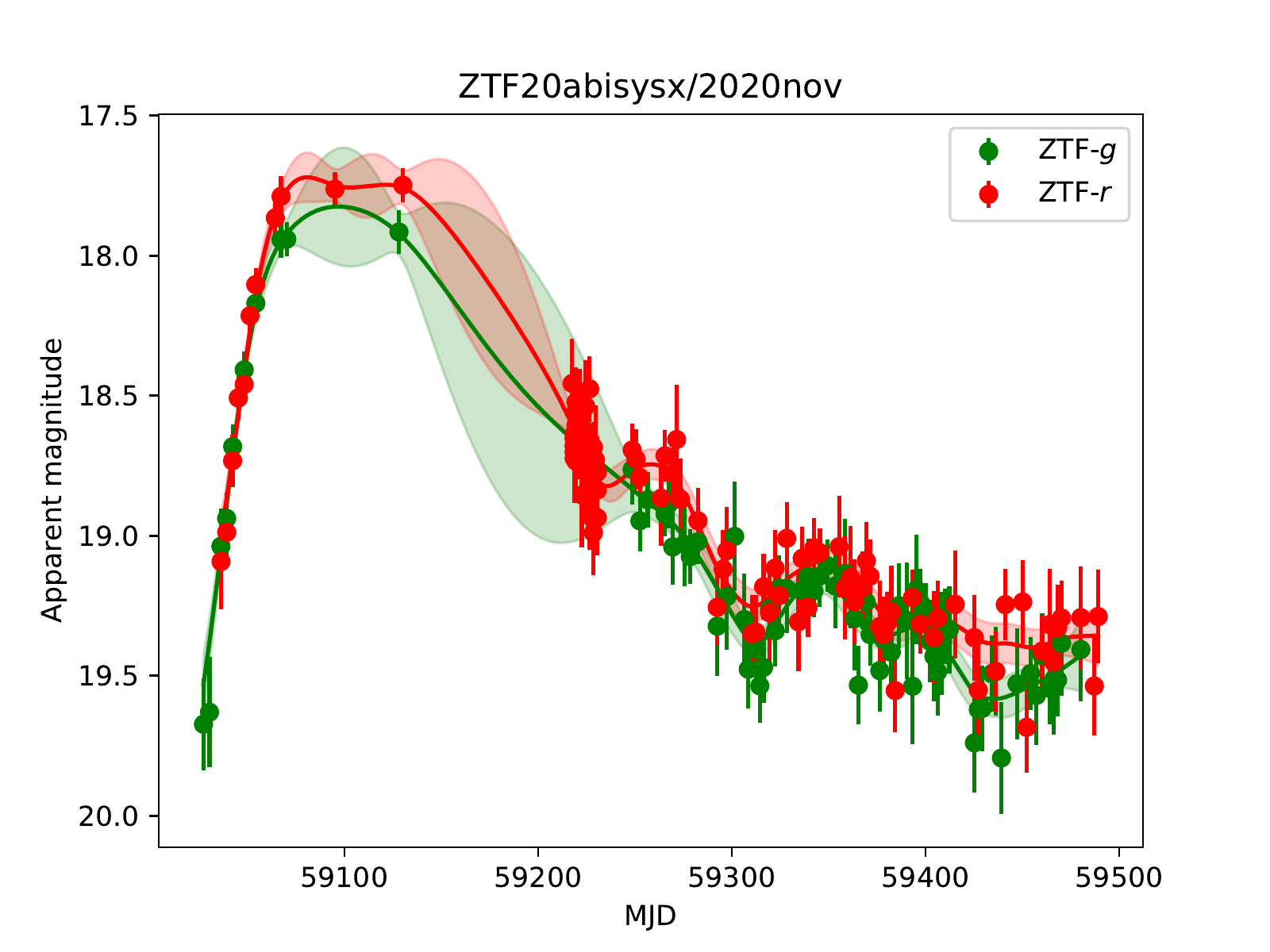}\hfill\includegraphics[width=80mm]{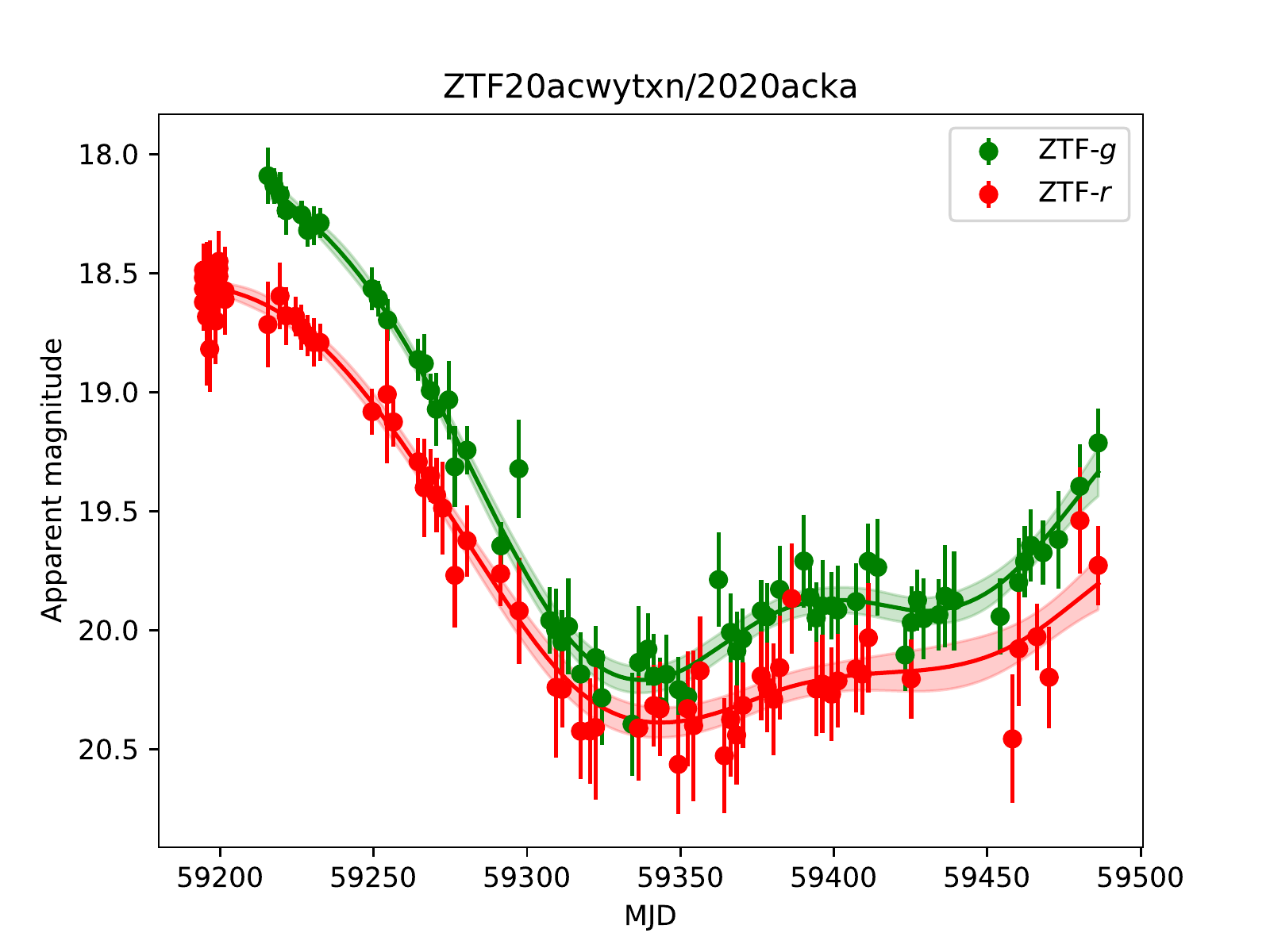}
    \includegraphics[width=80mm]{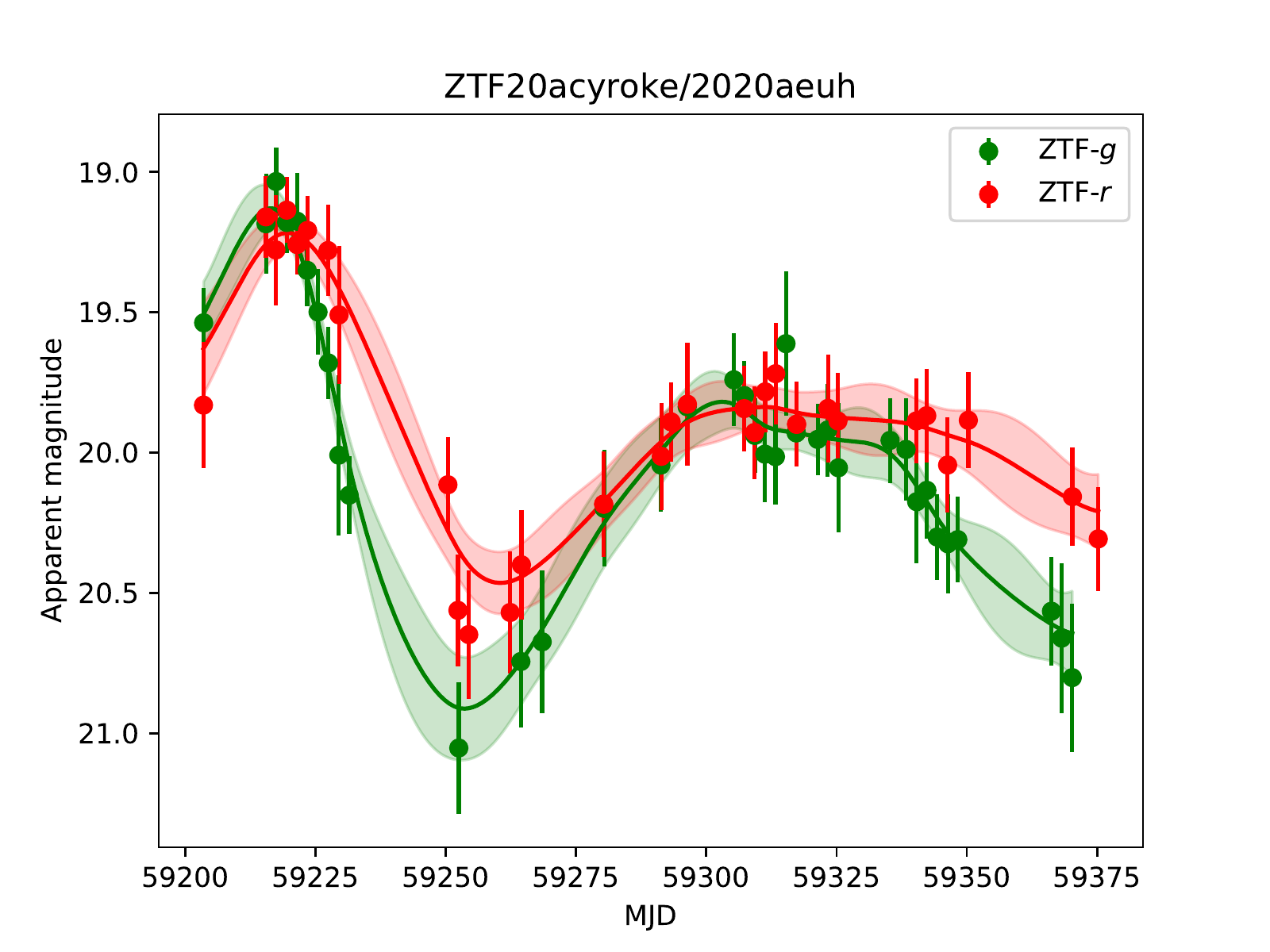}
    \caption{continued. The lines and symbols are the same as in Fig.~\ref{fig:exlcs}. 
    The sources in the the top panels have been typed as TDEs (ZTF20abisysx/2020nov \citealt{ Dahiwale2020TNSCR3800}; ZTF20acwytxn/2020acka \citealt{Hammerstein2021TNSCR.262}), while the source in the bottom panel is an SN~Ia \citep[2020aeuh][]{Weil2021TNSCR.812}.}
\end{figure*}

\subsection{Bumpy extragalactic transient categories}
\begin{itemize}
    \item {\it Core-collapse supernovae}: As can be seen in Table~\ref{tab:result}, the majority of the bumpy transients are core-collapse SNe, which include stripped-envelope (SE) SNe (SNe~IIb, SNe~Ib/c), SNe~IIn, and superluminous SNe (SLSNe; see \citealt{Filippenko-1997} and \citealt{Gal-Yam-2017} for reviews of SN classification). Some of these sources exhibit light curves with double peaks at early times ($\lesssim 100$~days), such as ZTF19aamsetj/2019cad, ZTF19abxjrge/2019ply, ZTF20adadlqm/2020adnx, ZTF20aamttiw/2020\-cpa, ZTF21abotose/2021ugl,  ZTF21aabxjqr/2021\-pb, ZTF21aaqvsvw/2021heh, and ZTF21abghxht\-/2021qbc. This kind of morphology is expected for SE~SNe, though it has also been seen in at least one SN~IIL 2007fz \citep{Faran-2014}, and the first peak is attributed to emission from the cooling envelope (see \citealt{Modjaz-2019}, and references therein). Indeed, the examples above are mostly SE~SNe, but also include a few with SN~II labels (e.g., 2020cpa).       
    
    Our results also include SNe~IIn and hydrogen-rich SLSNe (i.e., SLSNe-II). In particular, ZTF19abclykm/2019meh, ZTF19aceqlxc/2019smj, ZTF20aacbyec/2019zrk, and ZTF20acklcyp/2020xkx exhibit prominent bumps in their late-time light curves likely powered by CSM interaction \citep{Smith-2017}. Interestingly, 2019meh (a SLSN-II) has at least three peaks --- the third one currently rising (see Fig.~\ref{fig:bumps}), while 2020xkx (also a current event) shows two peaks, with the second one much broader ($\gtrsim 300$~days) and brighter than the first peak in the red passband. 
    
    The peak absolute magnitude of 2020xkx determined from the second peak in the red is at least $M_{r}=-19.02${\footnote{We do not account for host-galaxy extinction, since it is beyond the scope of this paper.}} (corrected for foreground extinction using the line-of-sight reddening from \citealt{Schlafly-2011}). Based on a spectrum taken a few days after discovery, it was classified as an SN~IIn \citep{Gromadzki-2020}. We obtained a spectrum of this event on 2021 August 12 (UT dates are used throughout this paper), 298~days after the first detection by ZTF, shown in Fig.~\ref{fig:spectra}. Emission lines of various species typical in such SNe are visible, prominent among which are Balmer and Paschen lines as well as the Ca~II near-IR triplet. A hint of [Ca~II] $\lambda\lambda7291$, 7324 is also present. The Balmer lines exhibit broad profiles (see Fig.~\ref{fig:spectra}) and ${\rm H}\alpha$ and ${\rm H}\beta$ in particular appear to develop double peaks, pointing to the SN ejecta interacting with CSM having a disk-like geometry \citep{Leonard-2000}. Its light curve, particularly the luminous peak combined with the broad shape, bears a striking resemblance to that of the peculiar iPTF14hls \citep{Arcavi-2017} and SN~2020faa \citep{Yang-2021}.  Continued follow-up observations of this event will be able to provide further constraints.

    Another extraordinary event in our sample is SLSN-II 2019meh with a peak absolute magnitude $M_{r}=-22.55$. Its multiple peaks are reminiscent of iPTF14hls; however, the light curve of 2019meh appears choppier. We obtained two epochs of  late-time spectra 803 and 859~days after its initial detection by ZTF, shown in Fig.~\ref{fig:spectra}. There is no dramatic evolution in the spectra between these two epochs. Strong emission lines of Balmer as well as Fe~II are visible. We also see the presence of [S~II] $\lambda\lambda6716$, 6731, prominent in H~II regions \citep[e.g.,][]{Peimbert-2017}, indicating that the SN explosion likely occurred in an H~II region, which is consistent with a massive-star progenitor given the type of this SN. The shape of the ${\rm H}\alpha$ profile (shown in the inset) closely resembles that of iPTF14hls seen in its late-time (+1153~d) spectrum studied by \citet{Andrews-2018}. These authors attributed the unique line profile to interaction with an asymmetric (disk-like) CSM and the multiple rebrightenings in its light curve (also seen for 2019meh) to multiple shells or variation in density of the CSM. These arguments likely apply to 2019meh as well, but a comprehensive analysis of earlier and late-time follow-up observations spanning multiple wavelengths will be required for conclusive evidence.                
    
    Detailed analyses of the spectra shown here are beyond the scope of this paper. The data can be accessed at the virtual storage space of Astro Data Lab {\url{soraisam://public/spectra_bumpySNe}} {\footnote{Astro Data Lab's \texttt{storeClient} library ({\url{https://datalab.noirlab.edu/docs/api/dl.html\#module-dl.storeClient}}) has to be used}}. 
    
    Our sample also contains one hydrogen-poor SLSN, ZTF20abobpcb/2020qlb (see Fig.~\ref{fig:bumpies}). 
    Recently, \citet{Hosseinzadeh-2021} established that bumps are rather prevalent in the post-peak light curves of 34 SLSNe-I they analyzed, finding both intrinsic (i.e., the central engine) and extrinsic (CSM interaction) factors as viable explanations. 2020qlb most likely belongs to this category of SLSNe-I.        
    
    Many of the SE-SNe and SNe~II in our sample also exhibit late-time ($>50\mbox{--}100$~days) rebrightening in their light curves (e.g., ZTF21abcjpnm/2021njo, ZTF20aadchdd/2020cd, ZTF21aanvncv/2021efd, ZTF20abxpoxd/2020sgf, ZTF19abgfuhh/2019lgc), which can be ascribed to CSM interaction (see \citealt{Sollerman-2020, Sollerman-2021}; one of the SE-SNe, 2019tsf, studied by these authors has also been selected by our algorithm).

\begin{figure*}
    \centering
    \includegraphics[width=180mm]{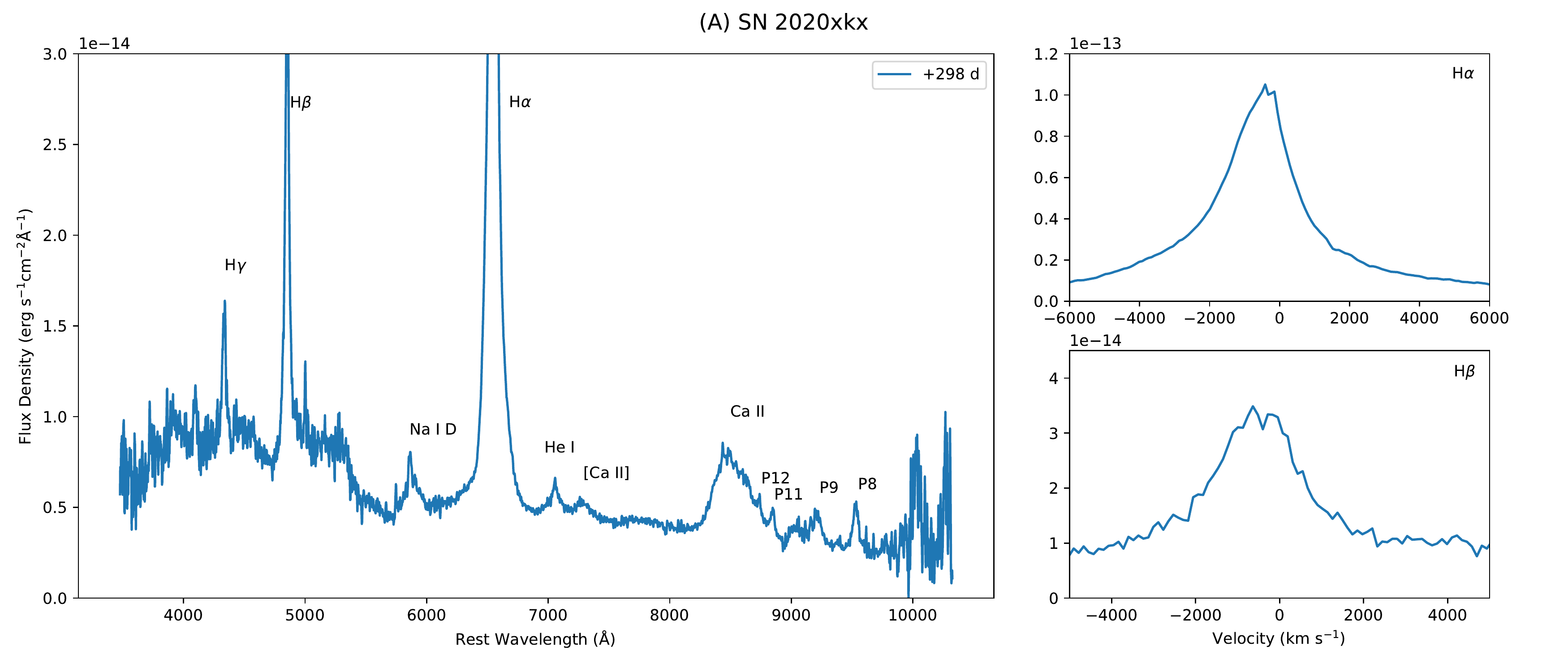}\\
    \includegraphics[width=160mm]{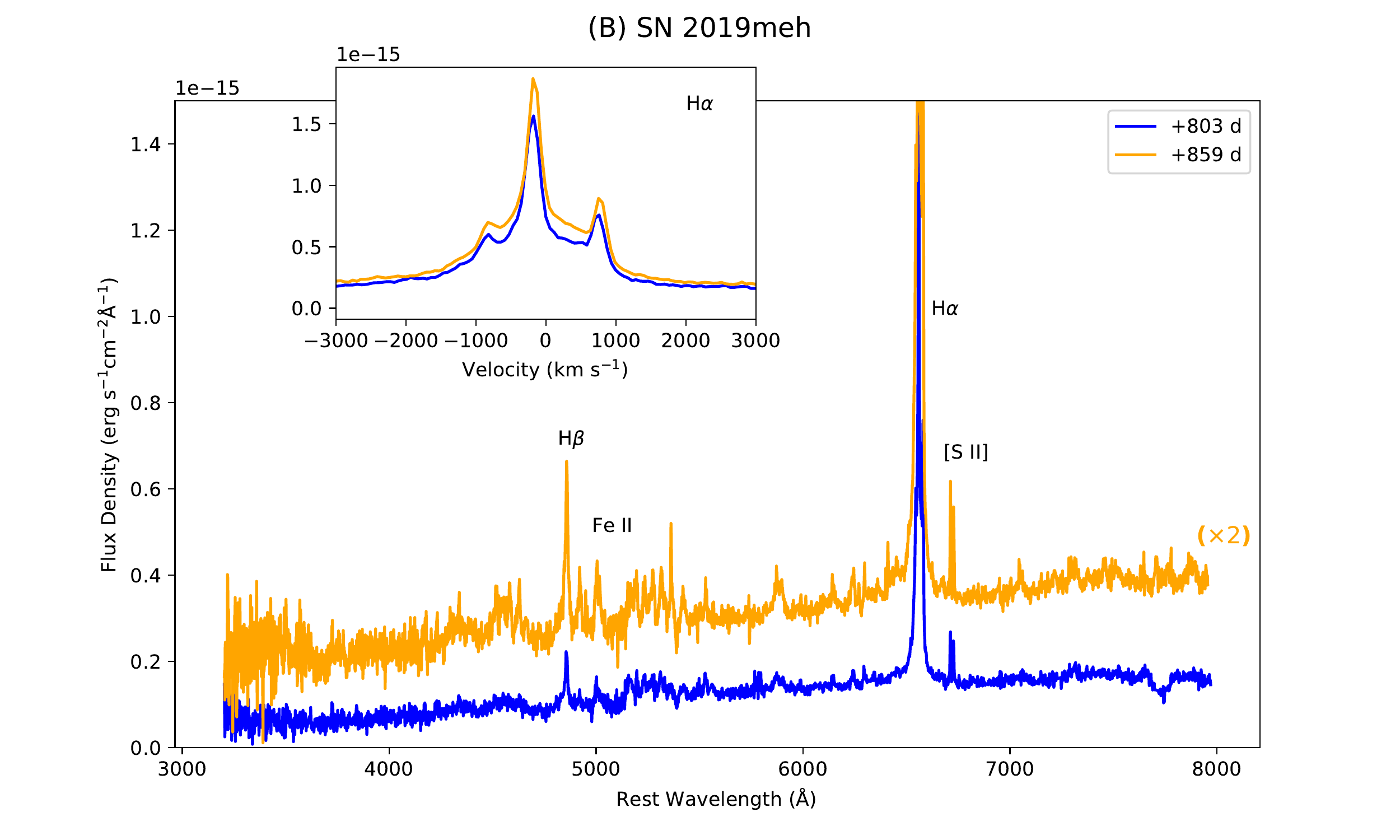}
    \caption{Late-time spectra of bumpy extragalactic transients obtained using the Kast double-beam spectrograph on the Shane 3\,m telescope at Lick Observatory. The epoch of the individual spectrum with respect to its first detection by ZTF is shown in the legend. The dip toward the red end (around 7800~\AA) at epoch +803~d for SN~2019meh is due to imperfect cosmic-ray removal.}
    \label{fig:spectra}
\end{figure*}

\begin{figure*}
    \ContinuedFloat
    \centering
    \includegraphics[width=160mm]{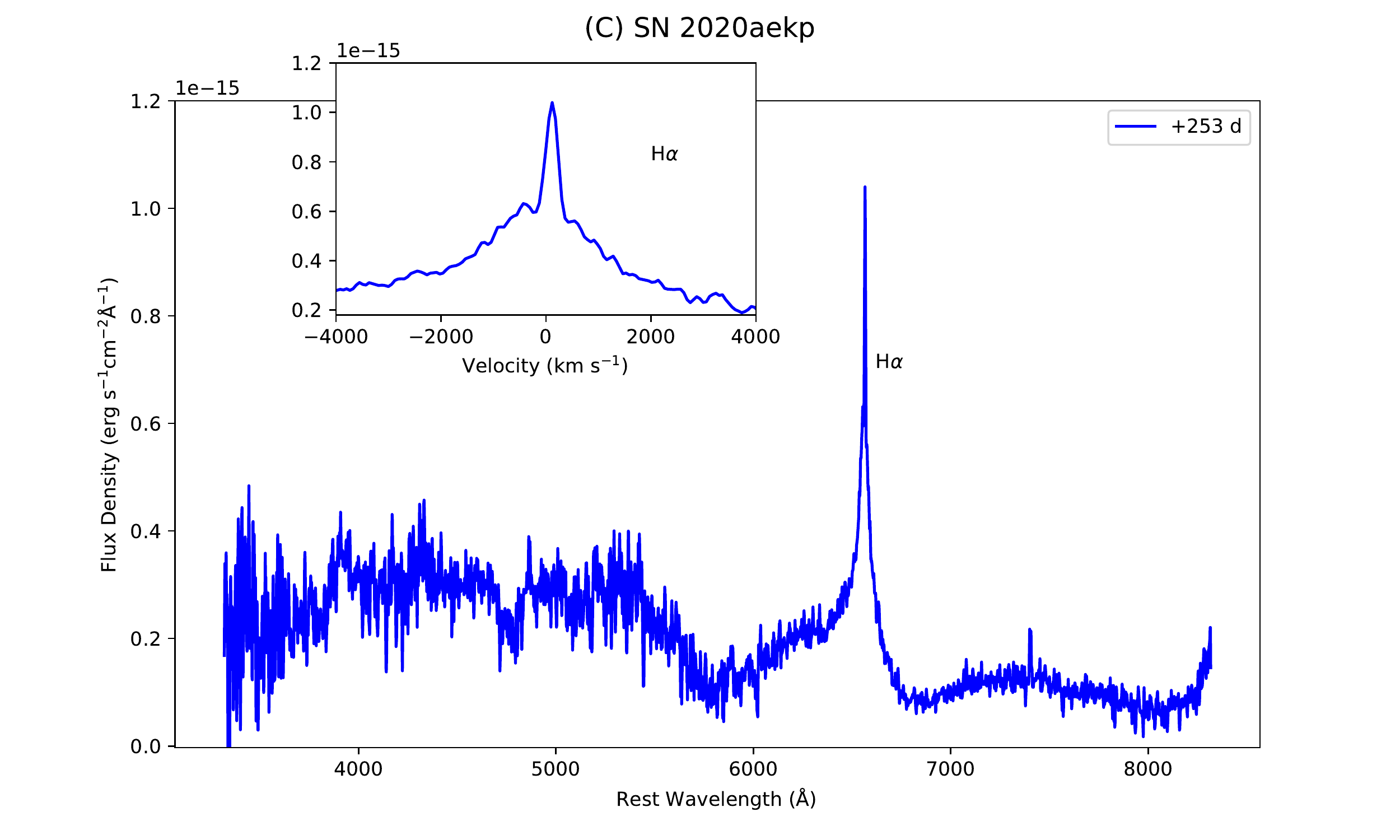}\\
    \includegraphics[width=160mm]{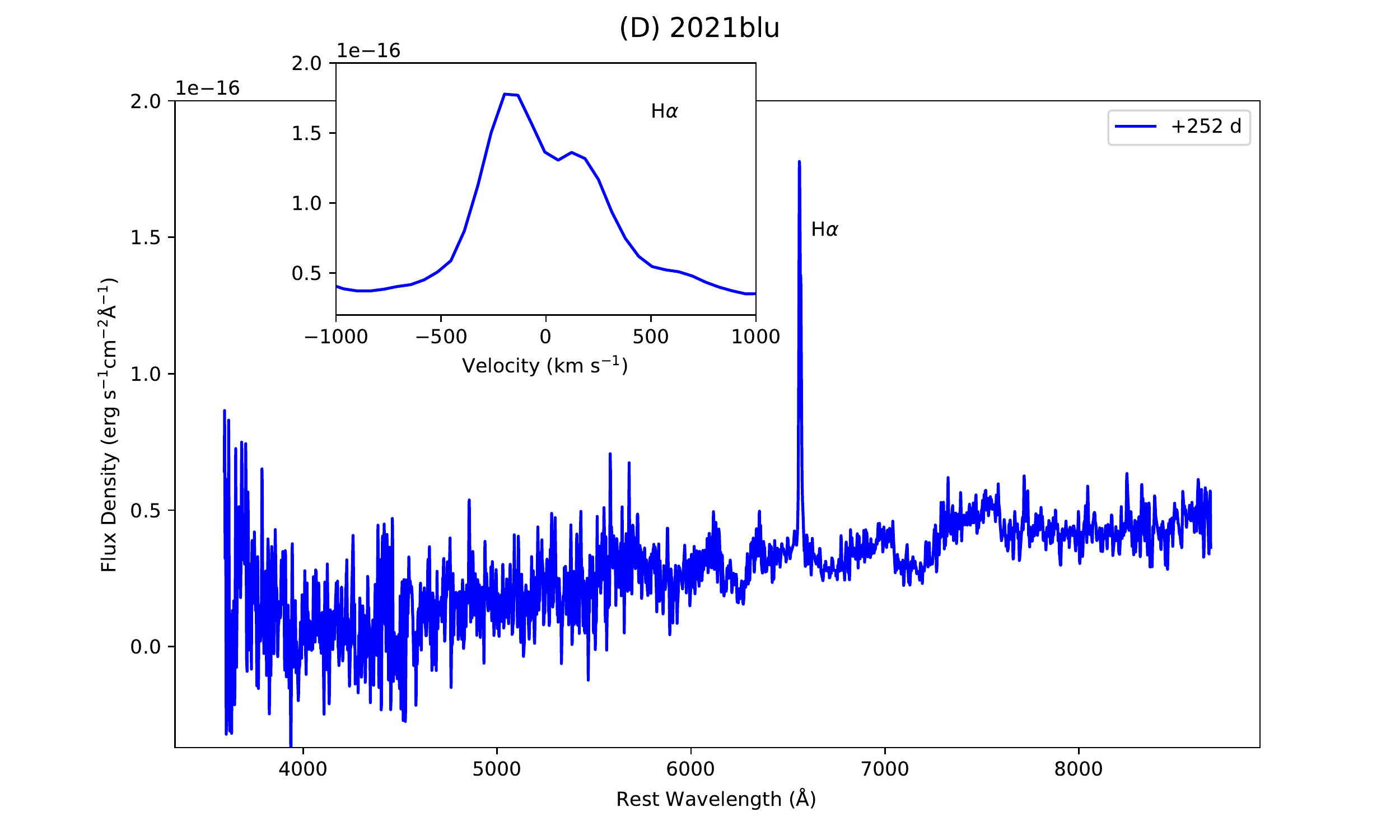}
    \caption{continued.}
\end{figure*}

    \item {\it Thermonuclear supernovae}: The subtype of thermonuclear SNe known as SN~Ia-CSM (also known as SNe~Ia/IIn or SN~2002ic-like) is characterized by signatures of interaction between the SN ejecta and CSM, which include overluminous, slowly decaying light curves, with bumps in some cases (e.g., the prototype SN~2002ic; \citealt{Wood-Vasey-2004}), and relatively narrow Balmer emission lines. We find a few bumpy transients that have been classified as SNe~Ia (and SN~Ia-CSM for ZTF21aaabwzx/2020aekp) in our final sample. CSM interaction is thus the likely driver for the bumps seen in their light curves (e.g., ZTF20acyroke/2020aeuh, 2020aekp, etc.). We also obtained a spectrum of 2020aekp +253~days after first detection by ZTF (Fig.~\ref{fig:spectra}). A clear sign of reduced flux on the red side of ${\rm H}\alpha$ is visible, which is typical of SNe~Ia-CSM \citep{Silverman-2013}.     
    
    \item {\it Tidal disruption events (TDEs)}: A few TDEs are also included in our sample --- ZTF20acwytxn/2020acka, ZTF20abisysx/2020nov, and  ZTF21aanxhjv/2021\-ehb. Bumps are prominent in the declining phase of the light curves for the latter two TDEs, while 2020acka exhibits a long monotonic rebrightening. TDEs are typically characterized by smoothly declining optical light curves, and the handful of discrepant examples such as PS18kh \citep{Holoien-2019} and 2018fyk/ASASSN–18ul \citep{Wevers-2019} tend to display atypical features like rebrightenings and plateaus only in their ultraviolet (UV) light curves, thought to be connected with processes in the accretion disk. The fluctuations in optical light curves shown by 2020nov and 2021ehb in our sample are perhaps similar to those seen in ASASSN-18jd, whose nature as either a TDE or a new class of nuclear transient is still unclear \citep{Neustadt-2020}. The rebrightening episode of 2020acka, though classified as a TDE by \citet{Hammerstein-2021}, resembles that of ZTF19abvgxrq/2019pev (also included in our sample; see below), which has been proposed as a member of the Trakhtenbrot AGN flares \citep{Trakhtenbrot-2019} by \cite{Frederick-2021}. These three events in our sample are currently active, and their continued late-time observations will help constrain their underlying physics.

    \item {\it Other}: Of the remaining transients in Table~\ref{tab:result}, ZTF21aagppzg/2021blu (see Fig.~\ref{fig:bumps}) is an LBV at redshift $z = 0.002$ in the Spider galaxy (UGC~5829). We obtained a spectrum of 2021blu with Lick Shane/Kast, 252~days after its first detection during the broad second bump, which is shown in Fig.~\ref{fig:spectra}. The spectrum is rather noisy; the object was quite faint ($\sim19.5$~mag) and the weather was poor.  However, a narrow emission line from ${\rm H}\alpha$ (${\rm FWHM}\approx650~{\rm km}~{\rm s}^{-1}$) can be clearly seen and hints of broad metal absorption lines.  2021blu is also an ongoing event and more late-time observations will be critical to establish the nature of the second outburst --- and whether it may be a terminal SN after all. 
    
    The unlabelled sources in Table~\ref{tab:result} include \texttt{ANTARES}-tagged nuclear transients  ZTF19abvgxrq/2019pev, ZTF19aaiqmgl/2019avd, and ZTF19aaciohh/2019baf. The first two have been studied by \citet{Frederick-2021} with additional follow-up data including spectroscopy and UV/X-ray observations. Based on spectroscopic similarity to AT~2017bgt, these authors have suggested 2019pev and 2019avd as members of the enhanced accretion-driven AGN flare class of \citet{Trakhtenbrot-2019}. 2019baf also likely belongs to this class given the similarity in its light curve evolution, even though spectropscopic observations are warranted for confirmation.     
    
\end{itemize}

\section{Conclusion} \label{sec:conclude}
We have presented a first systematic study of the archival alert database of \texttt{ANTARES} hosted on NOIRLab Astro Data Lab, focusing on extragalactic transient events for which more than 10,000 unique loci were found. Modeling their light curves with Gaussian Processes, we have narrowed down the sample size to just $\sim 1.4$\% with events showing multiple peaks or rebrightenings. From a rigorous visual vetting, we identify 37 transients with multiple prominent bumps in their light curves. These events largely fall into one of four categories: core-collapse SNe, SNe~Ia-CSM, TDE/AT~2017bgt-like, and LBV outbursts. Examples include luminous SNe with  iPTF14hls-like multiple peaks 2019meh and 2020xkx, and LBV 2021blu with a broad ($>100$~days) rebrightening episode, which may (or may not) be related to the terminal explosion of the star. Many of these events are currently ongoing, including all the notable events mentioned above, which make them ideal candidates for late-time follow-up observations to constrain their nature. For example, iPTF14hls has demonstrated the value of such observations at multiple wavelengths \citep{Sollerman-2019}. Its optical light curve shows a dramatic steep decline after $\sim 1000$~days, ruling out certain models such as fallback accretion and magnetars. X-ray monitoring of this object has only yielded upper limits, in contrast with other high-density CSM interaction events like SN~2010jl.

Our work demonstrates the value of mining archival time-domain data with added values from processing by the \texttt{ANTARES} alert-broker. With the upcoming Rubin alert data, we anticipate a richer scientific yield from \texttt{ANTARES} even from its archival database.

\begin{table*}[ht]
\caption{Extragalactic transients from ZTF with bumy light curves}\label{tab:result}
\renewcommand\arraystretch{1.0}
\renewcommand{\tabcolsep}{4.5pt}
\centering
\begin{tabularx}{1.0\textwidth}{llllll}
\hline
ZTF ID	
&RA (J2000)		
&DEC (J2000)
&IAU Name
&Type\tablenotemark{a}
&\texttt{ANTARES} link\\
\hline
ZTF18acrhegn & 200.871460 & 27.598350 & 2018jbd & -- & https://antares.noirlab.edu/loci/ANT2019r4lgk\\
ZTF19aaciohh & 268.000524 & 65.626643 & 2019baf & -- & https://antares.noirlab.edu/loci/ANT2019fi5vy\\
ZTF19aaduufr & 130.748509 & 59.567897 & 2019bjr & SN IIP & https://antares.noirlab.edu/loci/ANT2020mdnsy\\
ZTF19aaiqmgl & 125.903213 & 4.384020 & 2019avd & -- & https://antares.noirlab.edu/loci/ANT2020itnzy\\
ZTF19aatubsj & 257.278529 & 26.855685 & 2019fdr & SLSN-II & https://antares.noirlab.edu/loci/ANT2020a332y\\
ZTF19aamsetj & 137.179054 & 44.812814 & 2019cad & SN Ic & https://antares.noirlab.edu/loci/ANT2019vndjc\\
ZTF19aarfkch & 221.131639 & 70.455985 & 2019dwf & -- & https://antares.noirlab.edu/loci/ANT2019pjqws\\
ZTF19abclykm & 321.822735 & 64.416439 & 2019meh & SLSN-II & https://antares.noirlab.edu/loci/ANT2020agsa4\\
ZTF19abgfuhh & 3.725184 & 36.300906 & 2019lgc & SN IIb & https://antares.noirlab.edu/loci/ANT2019lpafi\\
ZTF19abvgxrq & 67.344703 & 0.618760 & 2019pev & -- & https://antares.noirlab.edu/loci/ANT2020pfqji\\
ZTF19abxjrge & 44.945755 & 15.296852 & 2019ply & SN IIb & https://antares.noirlab.edu/loci/ANT2020unoti\\
ZTF19aceqlxc & 117.419663 & 5.074203 & 2019smj & SN IIn & https://antares.noirlab.edu/loci/ANT2020edwny\\
ZTF19ackjszs & 167.136636 & -10.481770 & 2019tsf & SN Ib & https://antares.noirlab.edu/loci/ANT2020hnhs6\\
ZTF20aacbyec & 174.947491 & 19.929643 & 2019zrk & SN IIn & https://antares.noirlab.edu/loci/ANT2020bq46c\\
ZTF20aadchdd & 203.234621 & 54.042149 & 2020cd & SN II & https://antares.noirlab.edu/loci/ANT2020alxje\\
ZTF20aamttiw & 261.657181 & 38.208176 & 2020cpa & SN II & https://antares.noirlab.edu/loci/ANT2020mosa\\
ZTF20aarvuvj & 265.952909 & 22.426205 & 2020dvd & -- & https://antares.noirlab.edu/loci/ANT2020amjjs\\
ZTF20aatxryt & 167.946571 & 29.385115 & 2020eyj & SN Ia & https://antares.noirlab.edu/loci/ANT2020as3vu\\
ZTF20abisysx & 254.554098 & 2.117527 & 2020nov & TDE & https://antares.noirlab.edu/loci/ANT2020dxbmq\\
ZTF20abobpcb & 286.956514 & 62.963752 & 2020qlb & SLSN-I & https://antares.noirlab.edu/loci/ANT2020nzw4s\\
ZTF20abxpoxd & 50.719846 & 42.556106 & 2020sgf & SN Ic & https://antares.noirlab.edu/loci/ANT202024yj2\\
ZTF20acihinn & 340.694664 & 51.505972 & 2020vzz & -- & https://antares.noirlab.edu/loci/ANT2020aefccmq\\
ZTF20acklcyp & 350.117382 & 22.986906 & 2020xkx & SN IIn & https://antares.noirlab.edu/loci/ANT2020aekfzhy\\
ZTF20acwytxn & 238.758128 & 16.304531 & 2020acka & TDE & https://antares.noirlab.edu/loci/ANT2020afjo7na\\
ZTF20acxqaof & 84.822243 & 63.641719 & 2020uly & SN Ia & https://antares.noirlab.edu/loci/ANT20207mb5a\\
ZTF20acyroke & 188.966056 & 37.703481 & 2020aeuh & SN Ia & https://antares.noirlab.edu/loci/ANT2020afmdwmy\\
ZTF20adadlqm & 46.481311 & 41.608693 & 2020adnx & SN II & https://antares.noirlab.edu/loci/ANT2020afo32ha\\
ZTF21aaabwzx & 235.797447 & 17.813116 & 2020aekp & SN Ia-CSM & https://antares.noirlab.edu/loci/ANT2021mqlq\\
ZTF21aabxjqr & 146.194990 & 51.687393 & 2021pb & SN IIb & https://antares.noirlab.edu/loci/ANT2021bbfmu\\
ZTF21aagppzg & 160.643112 & 34.437380 & 2021blu & LBV & https://antares.noirlab.edu/loci/ANT2021djjs2\\
ZTF21aaqvsvw & 119.947061 & 25.355832 & 2021heh & SN IIb & https://antares.noirlab.edu/loci/ANT2021ixxiq\\
ZTF21aanvncv & 171.102978 & 23.649608 & 2021efd & SN Ib/c & https://antares.noirlab.edu/loci/ANT2021hgwrm\\
ZTF21aanxhjv & 46.949254 & 40.311343 & 2021ehb & TDE & https://antares.noirlab.edu/loci/ANT2021hhxq6\\
ZTF21abcjpnm & 241.779105 & 47.443639 & 2021njo & SN II & https://antares.noirlab.edu/loci/ANT2021n6cbs\\
ZTF21abfjlxb & 218.342847 & 19.128498 & 2021pkd & SN IIb & https://antares.noirlab.edu/loci/ANT2021pt3zq\\
ZTF21abghxht & 292.510897 & 35.046311 & 2021qbc & SN II-pec & https://antares.noirlab.edu/loci/ANT2021qc32w\\
ZTF21abotose & 246.982038 & 20.253160 & 2021ugl & SN IIb & https://antares.noirlab.edu/loci/ANT2021uuv34\\

\hline
\end{tabularx}
\begin{flushleft}
\tablenotetext{a}{Classification, if available, is obtained from the Transient Name Server (\url{https://www.wis-tns.org/}). The redshift information for these events can be seen using the \texttt{ANTARES} link.}
\end{flushleft}
\end{table*}

\begin{acknowledgments}
{\it Acknowledgments}: The ANTARES team gratefully acknowledges financial support from the National Science Foundation (NSF) through a cooperative agreement with the Association of Universities for Research in Astronomy (AURA) for the operation of the NSF's National Optical-Infrared Astronomy Research Laboratory, through an NSF INSPIRE grant to the University of Arizona (CISE AST-1344024, PI:~R.~Snodgrass), and through a grant from the Heising-Simons Foundation (2018-0909, PI:~T.~Matheson).
A.V.F.'s group at U.C. Berkeley is grateful for funding from the TABASGO Foundation, the Miller Institute for Basic Research in Science (where he is a Miller Senior Fellow), the Christopher J. Redlich Fund, and many individual donors. Research at Lick Observatory is partially supported by a generous gift from Google. This research uses services or data provided by the Astro Data Lab at NSF's National Optical-Infrared Astronomy Research Laboratory.       
\end{acknowledgments}


\software{\texttt{numpy} \citep{numpy}, \texttt{scipy} \citep{scipy}, \texttt{astropy} \citep{astropy}, \texttt{matplotlib} \citep{matplotlib}}

\bibliography{ref}{}
\bibliographystyle{aasjournal}


\clearpage
\appendix
\section{Extragalactic transients with bumpy light curves}\label{appendix:bumpies}

\begin{figure*}[h]
    \centering
    
    \includegraphics[width=80mm]{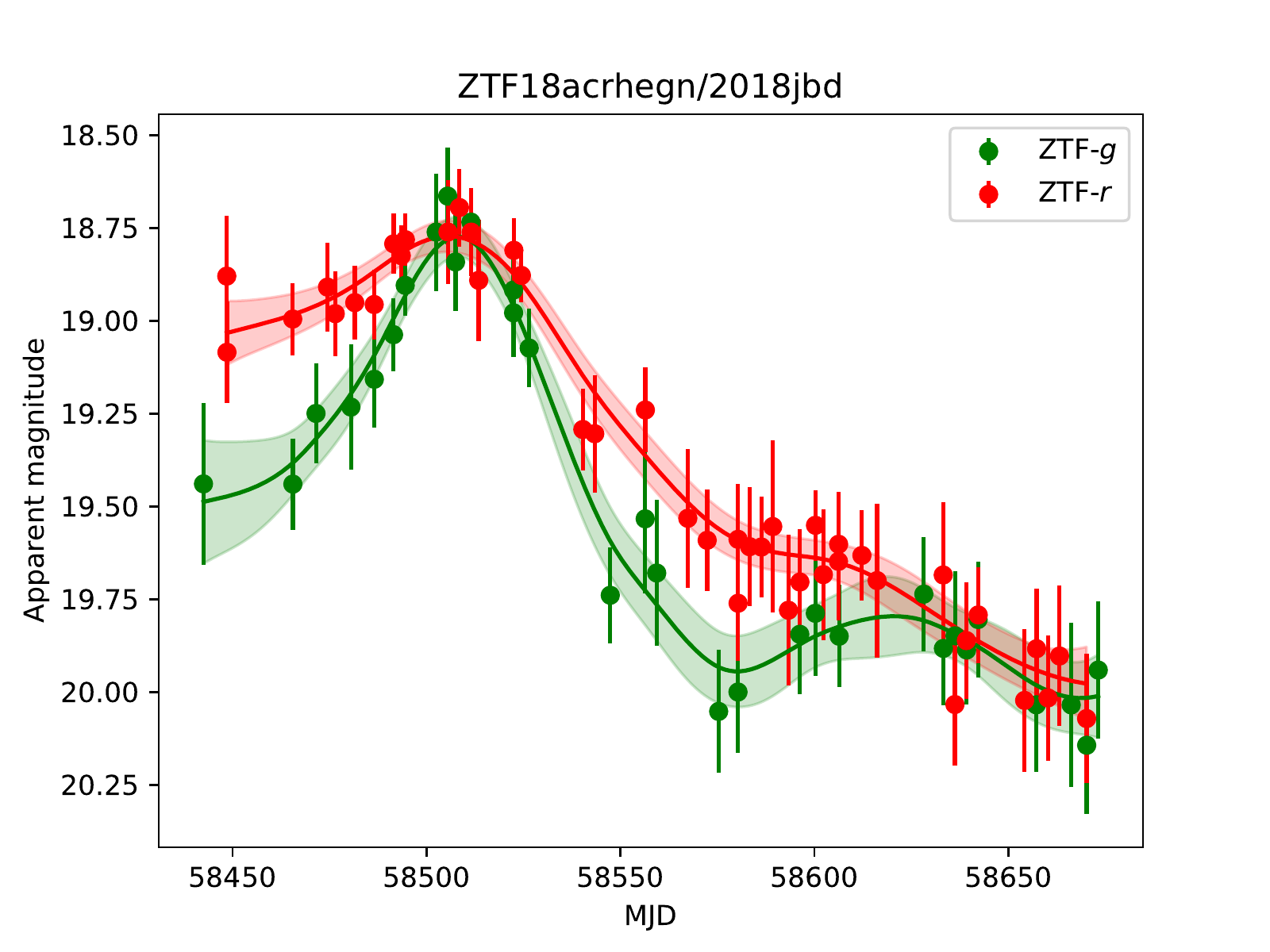}\hfill\includegraphics[width=80mm]{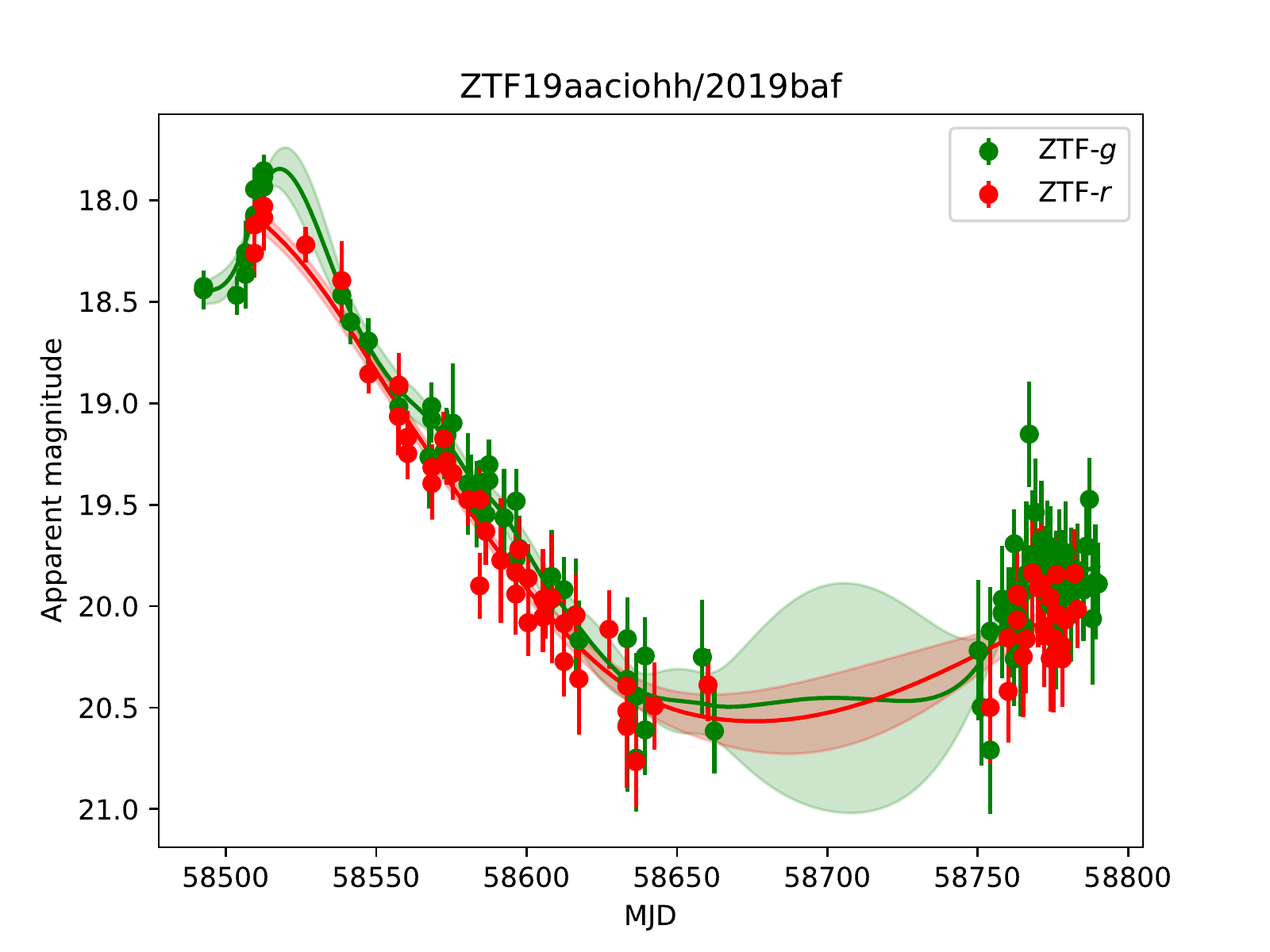}
    \includegraphics[width=80mm]{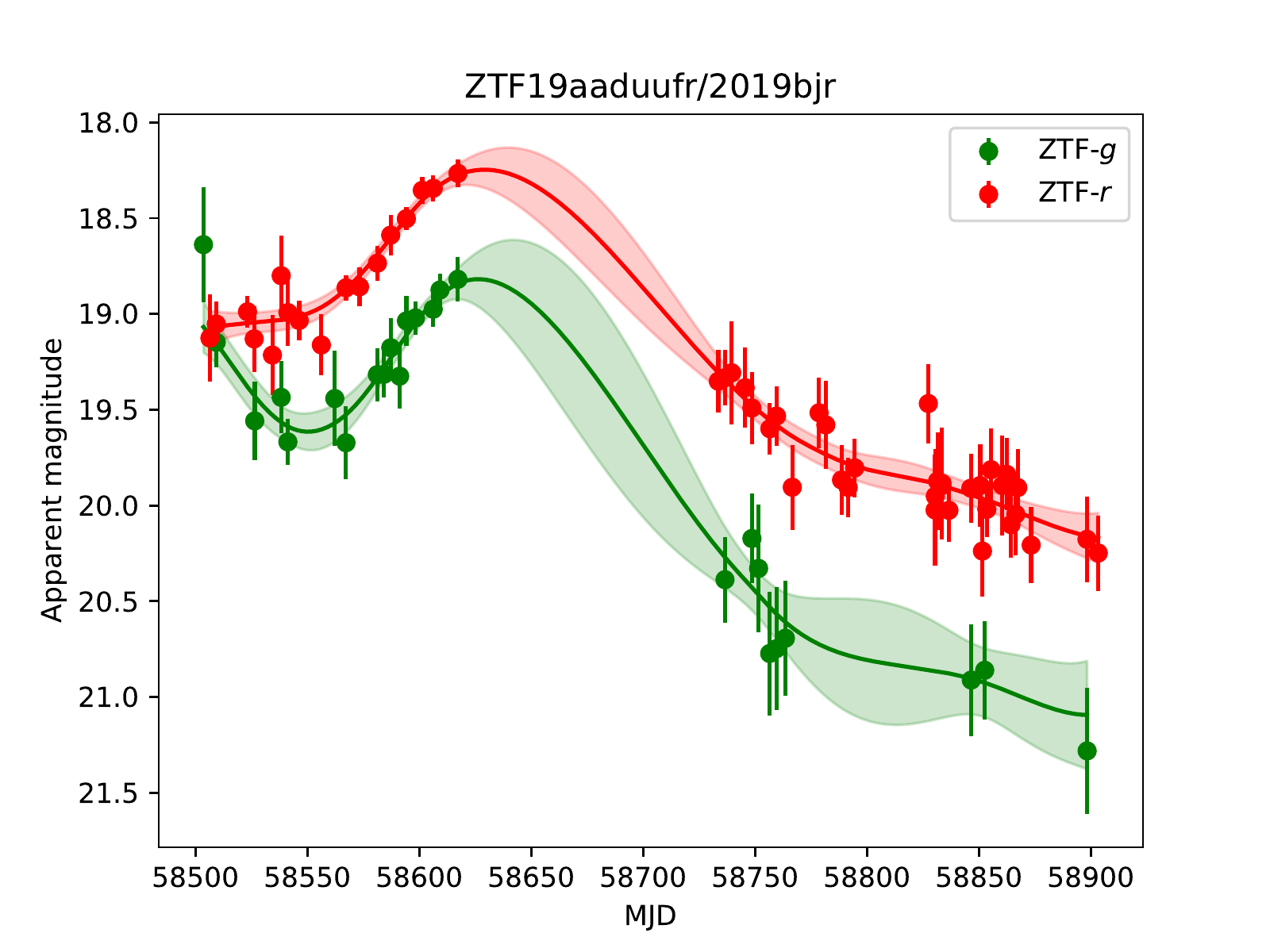}\hfill\includegraphics[width=80mm]{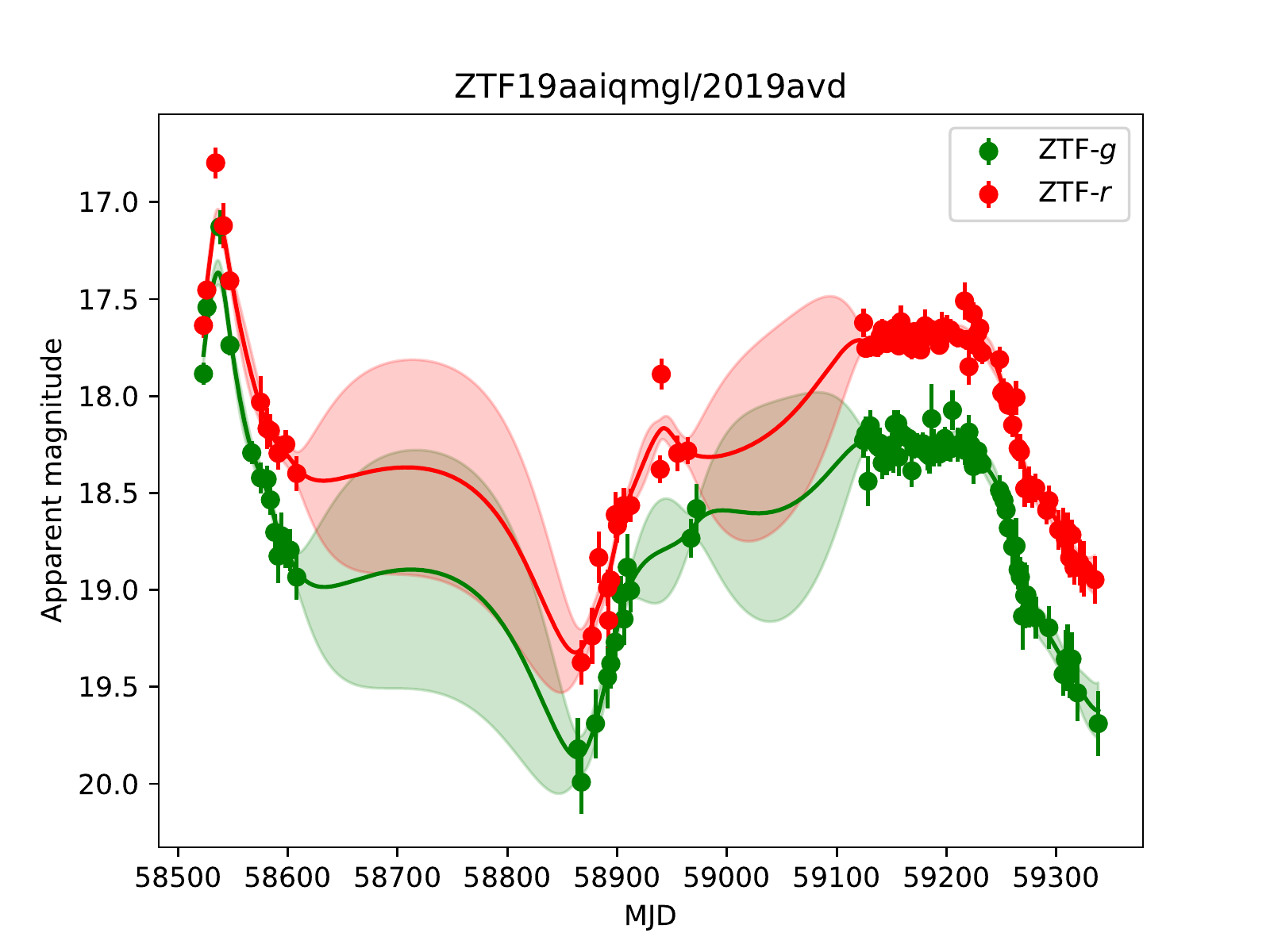}
    \includegraphics[width=80mm]{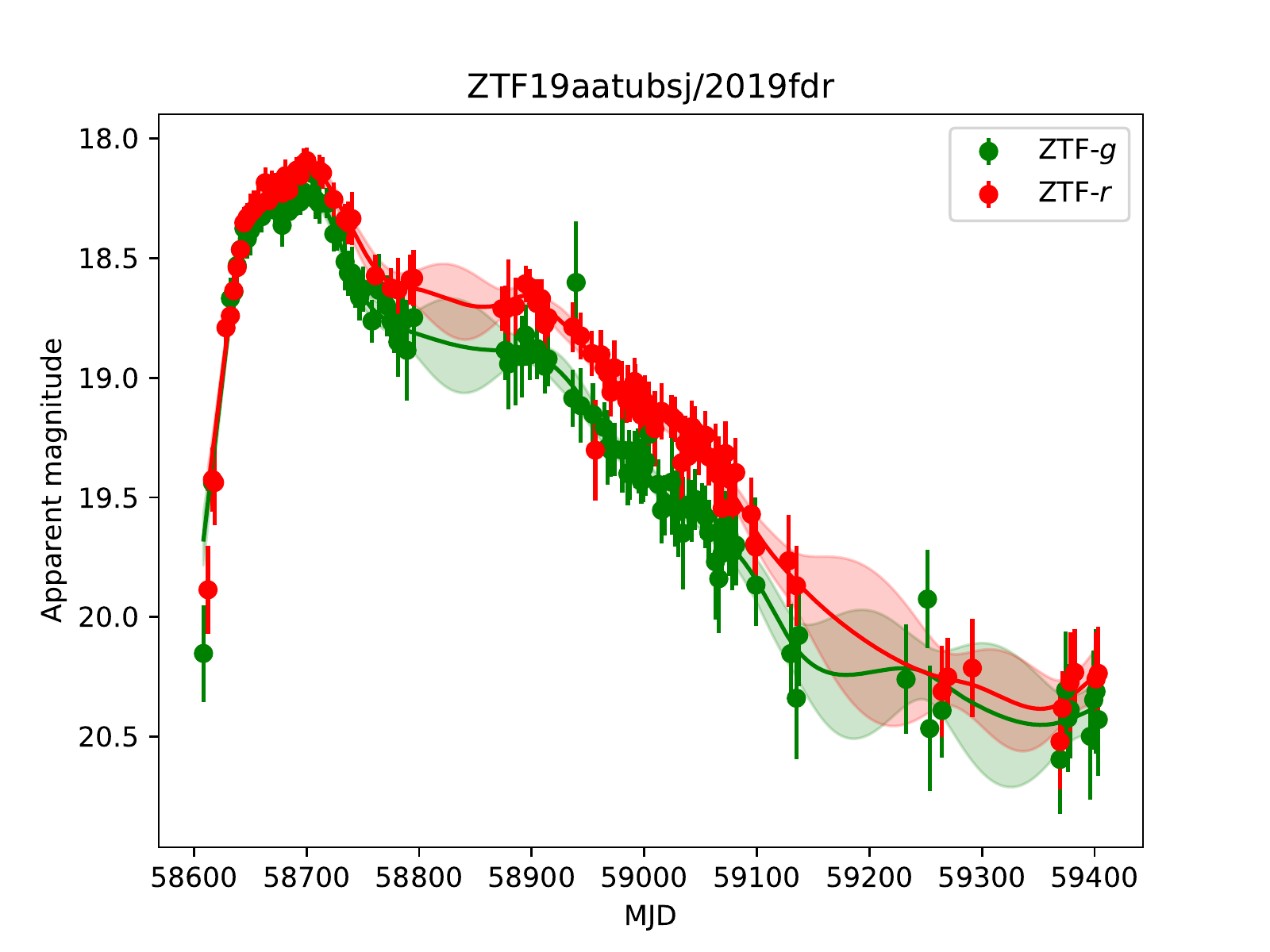}\hfill\includegraphics[width=80mm]{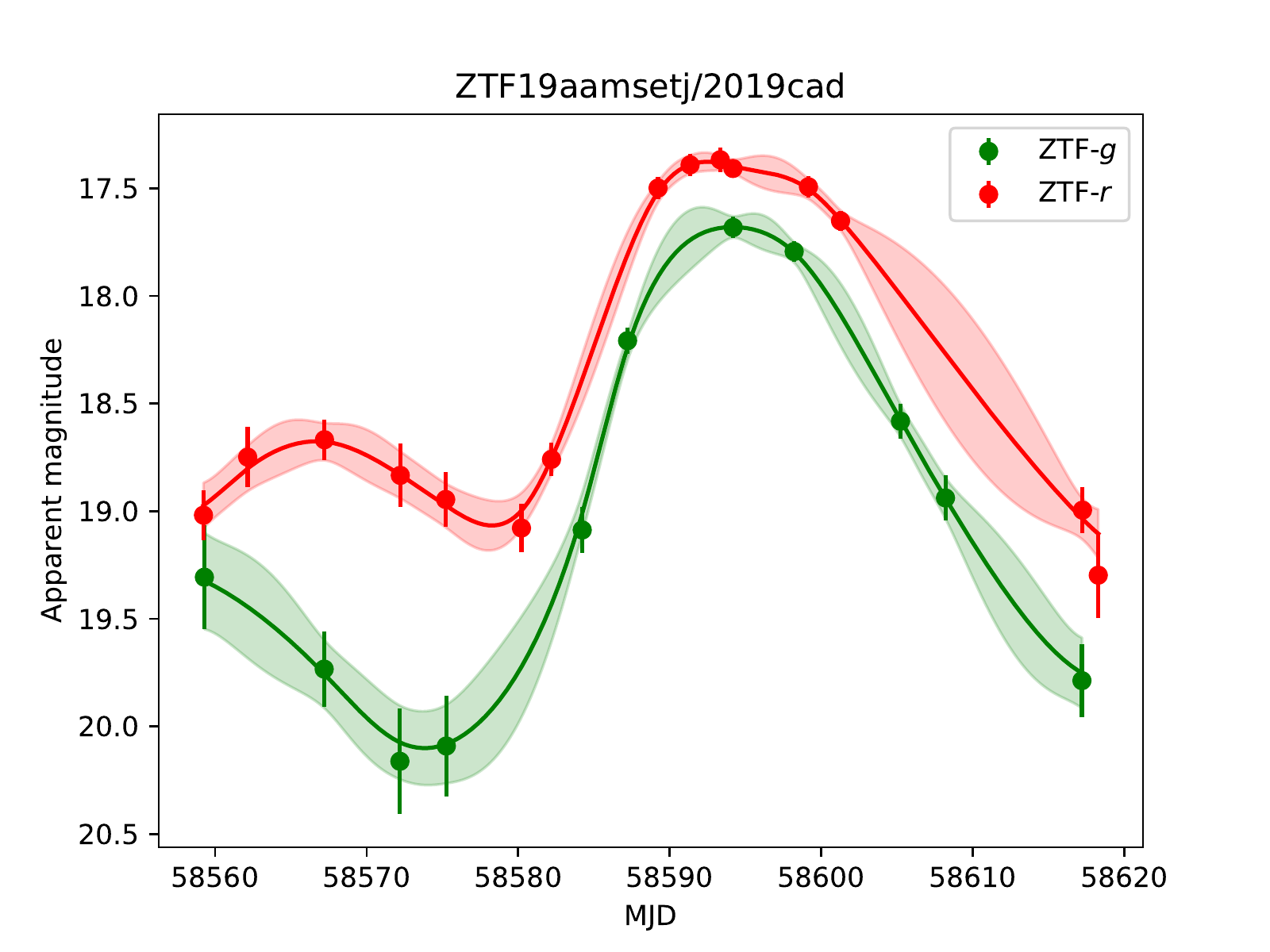}
    \caption{ZTF light curves of bumpy extragalactic transients. The lines and symbols are similar to those in Fig.~\ref{fig:exlcs}. Details of the individual sources are given in Table~\ref{tab:result}.}
    \label{fig:bumpies}
\end{figure*}

\begin{figure*}
    \ContinuedFloat
    \centering
    \includegraphics[width=80mm]{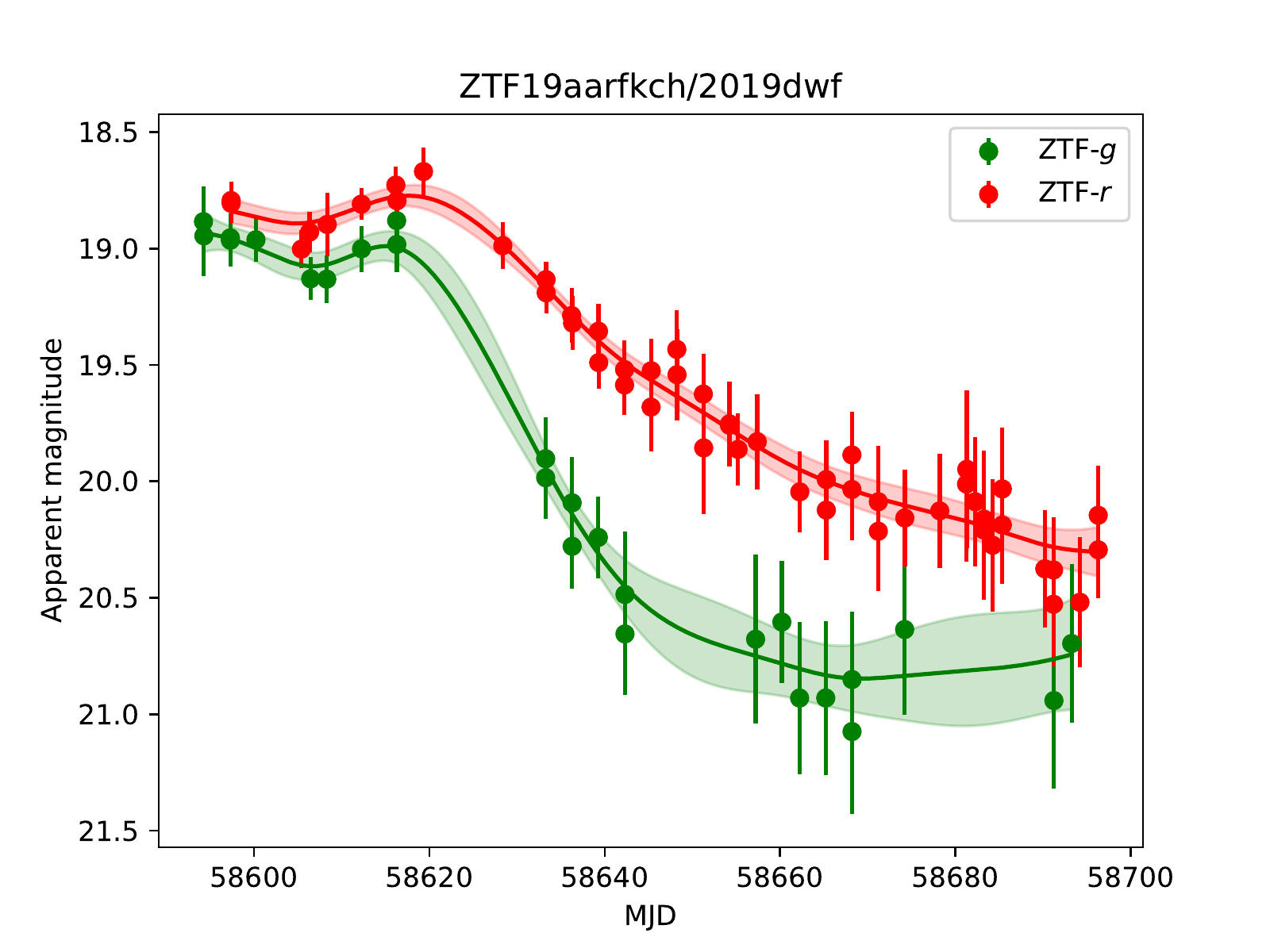}\hfill\includegraphics[width=80mm]{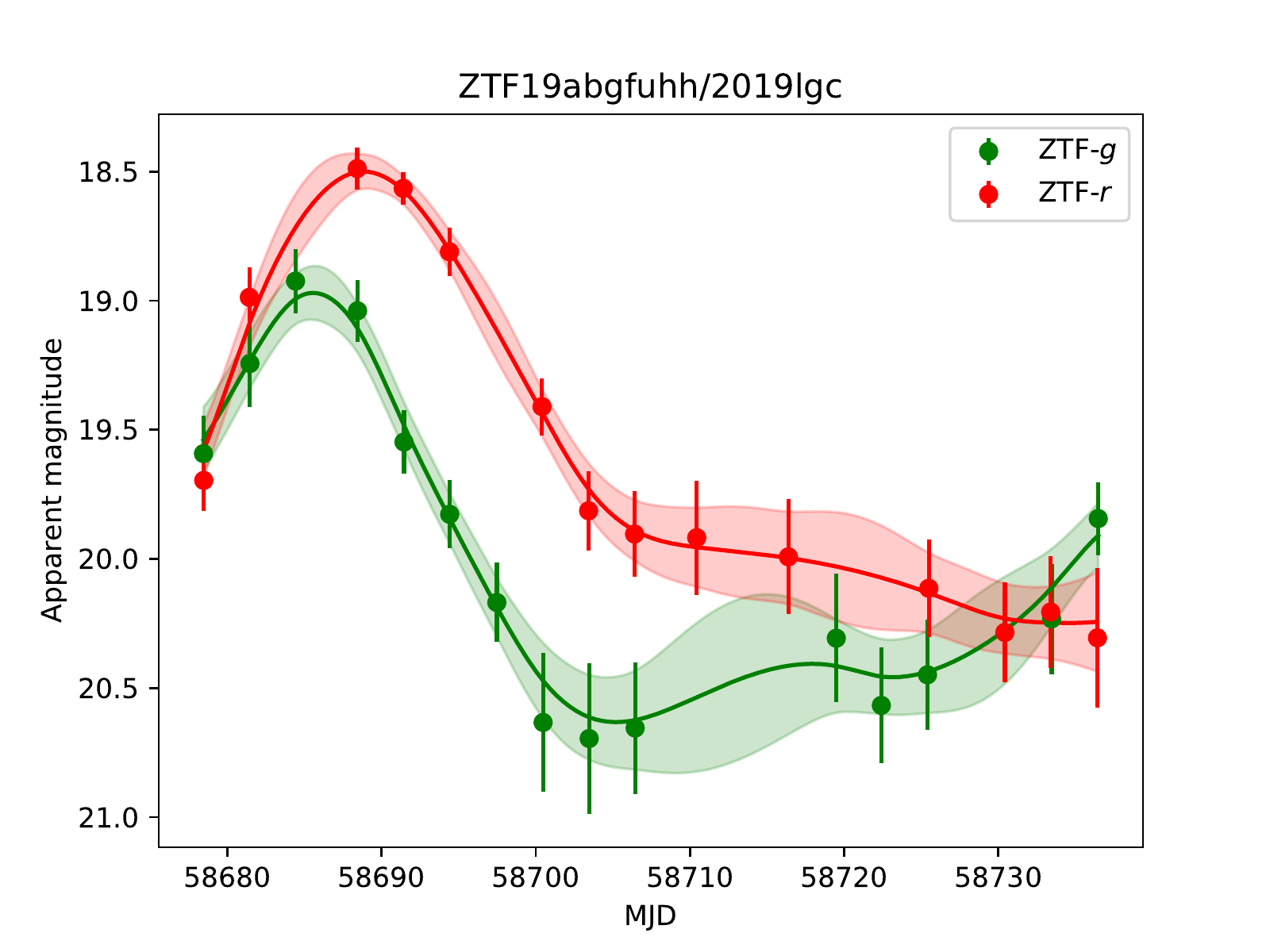}
    \includegraphics[width=80mm]{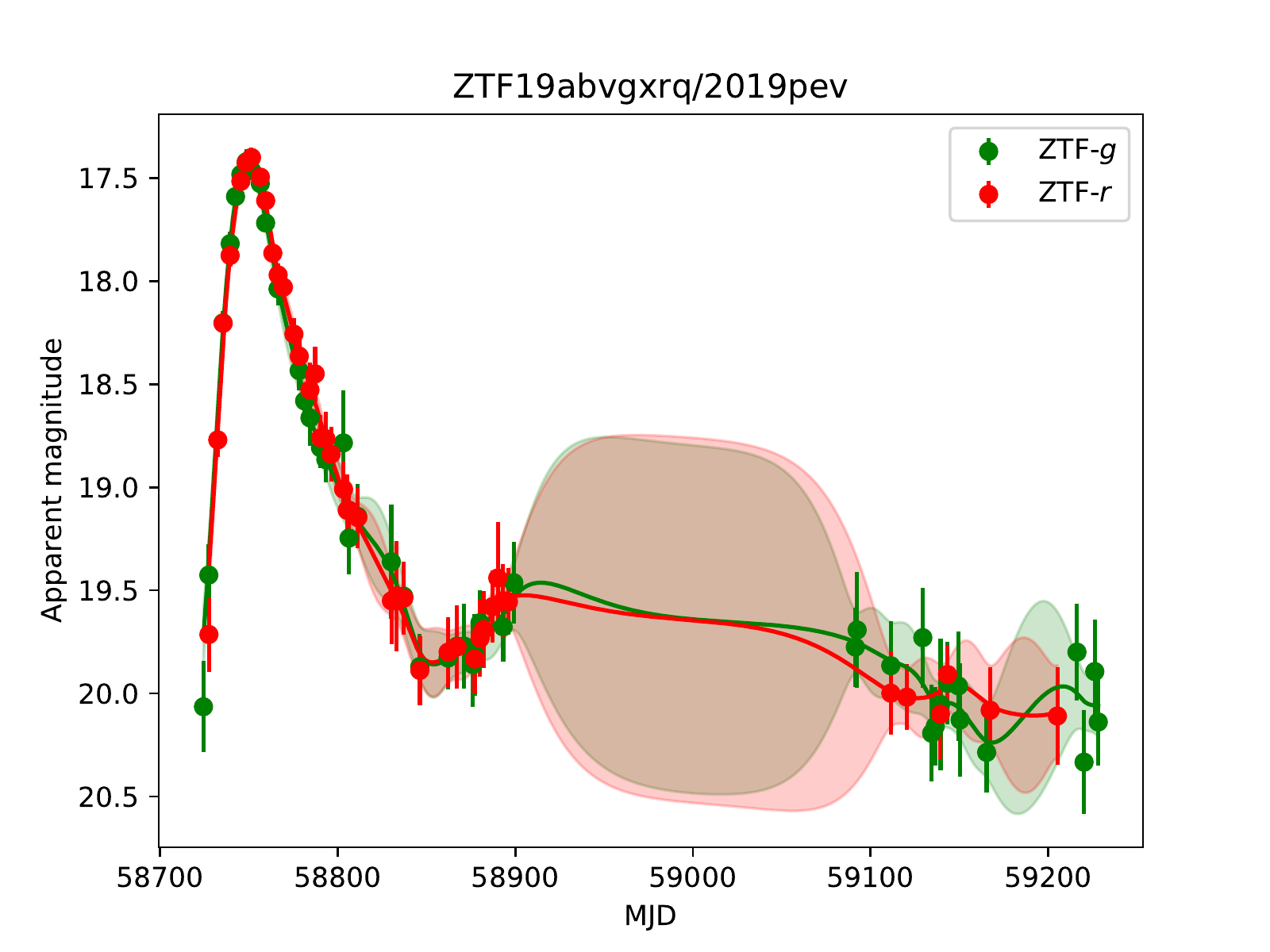}\hfill\includegraphics[width=80mm]{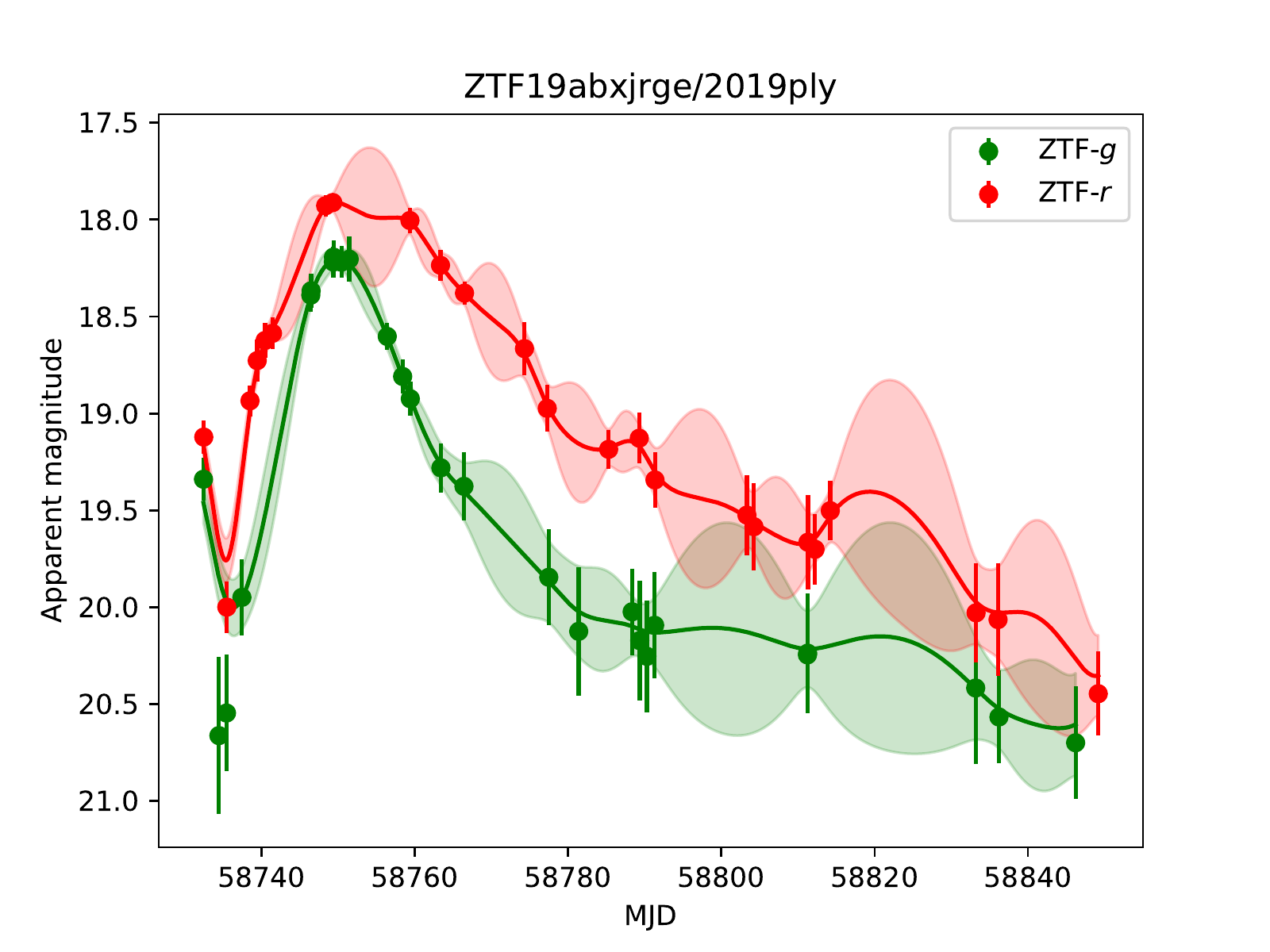}
    \includegraphics[width=80mm]{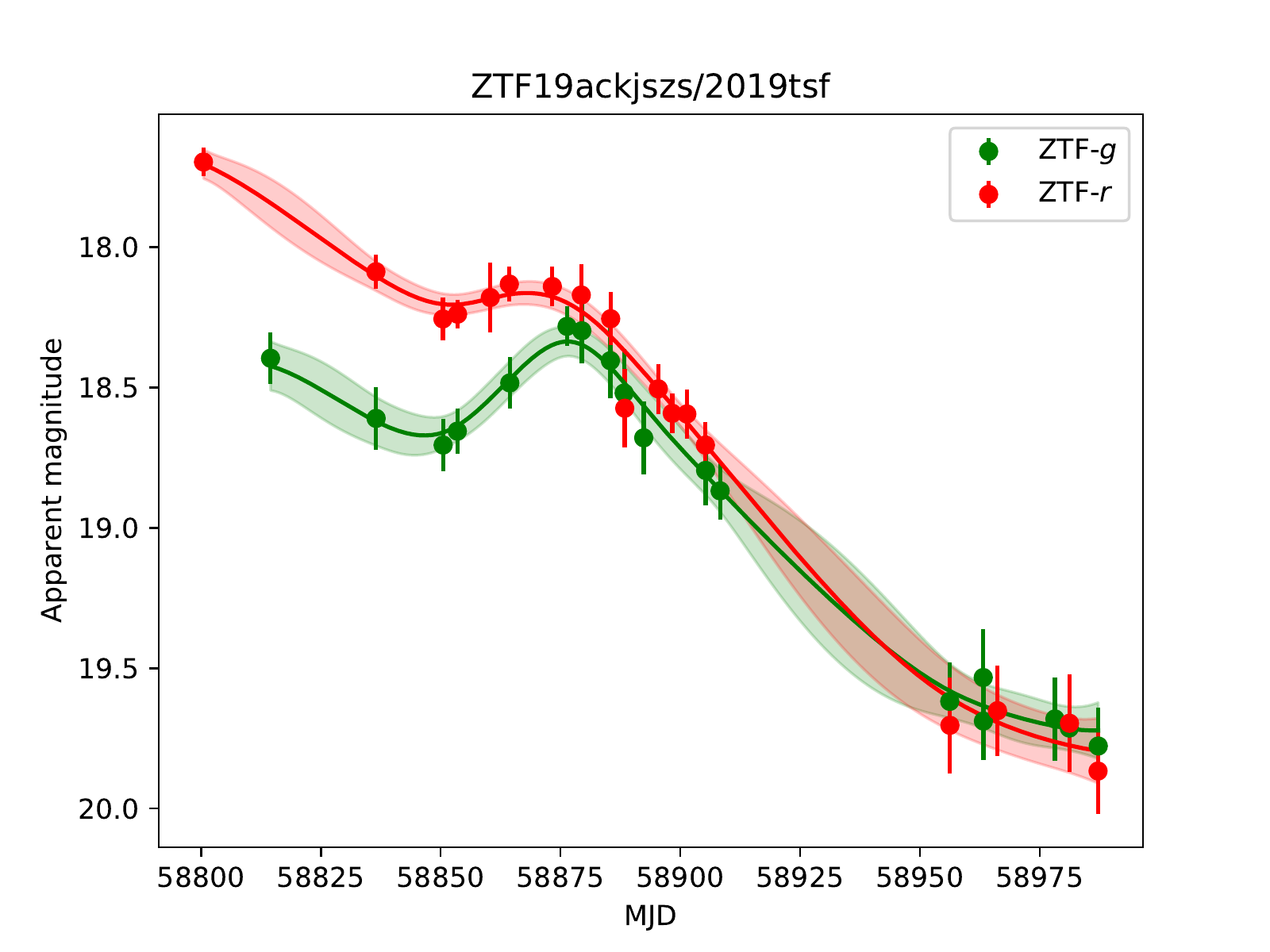}\hfill\includegraphics[width=80mm]{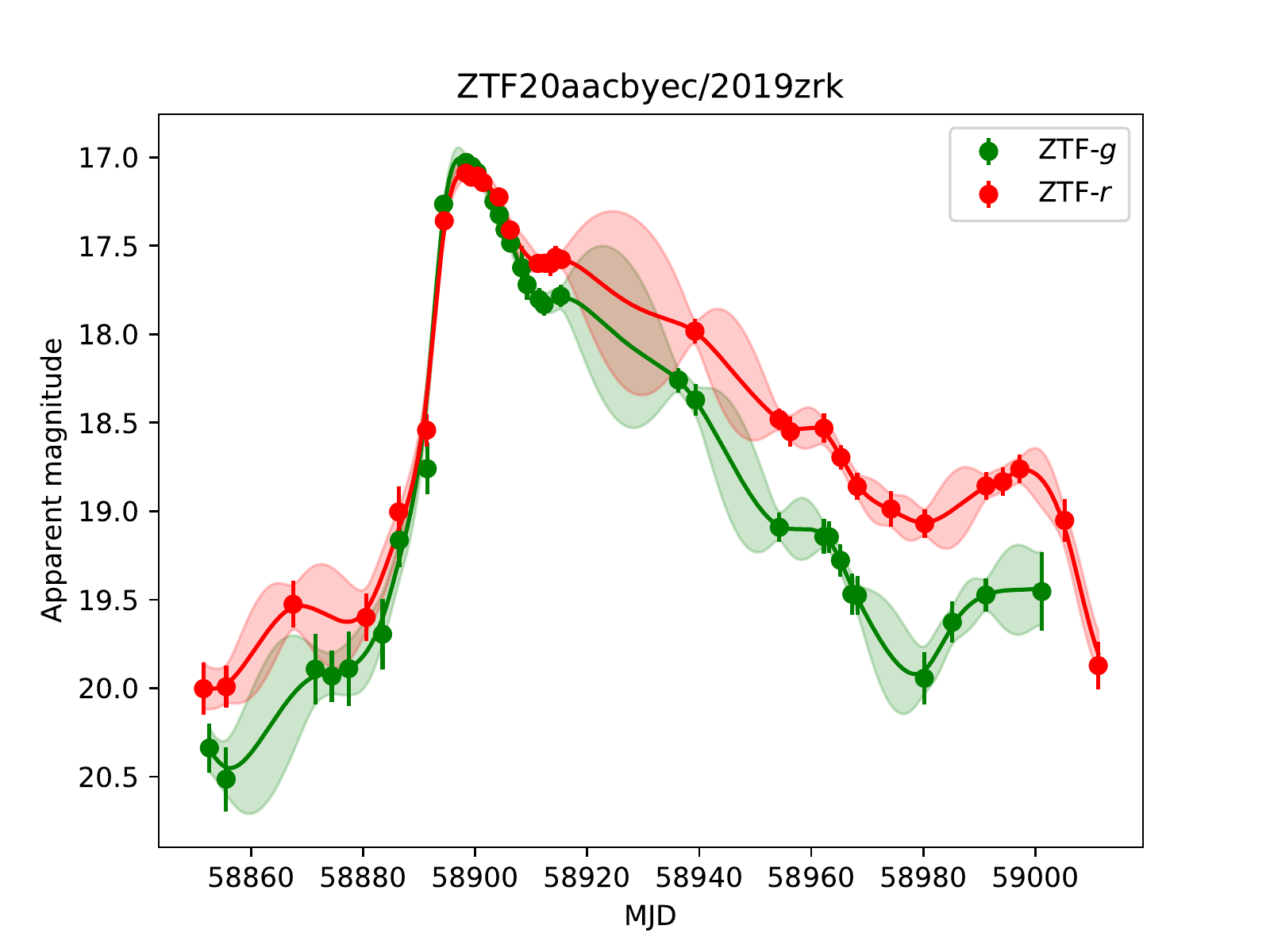}
    \caption{continued}
\end{figure*}

\begin{figure*}
    \ContinuedFloat
    \centering
    \includegraphics[width=80mm]{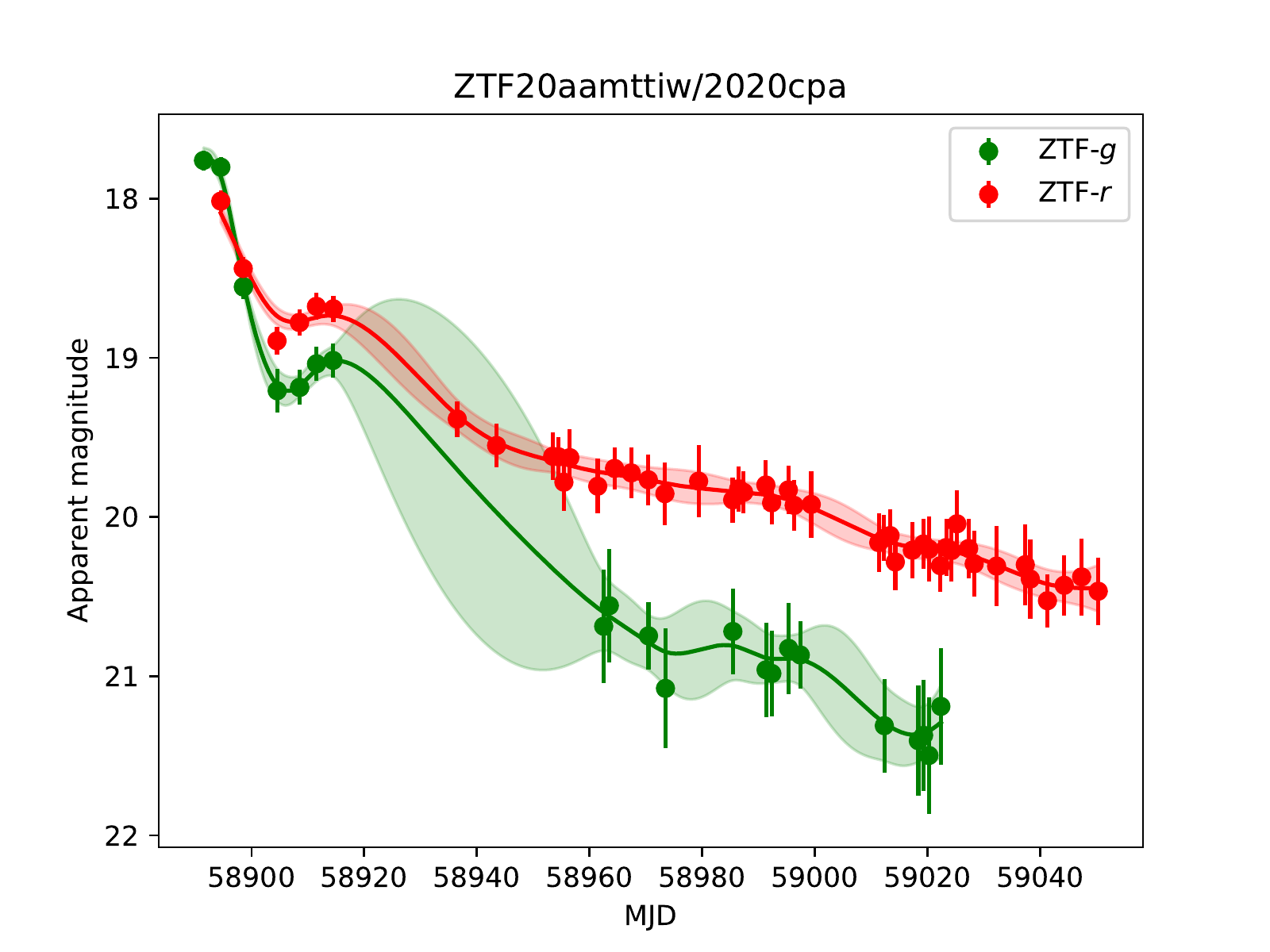}\hfill\includegraphics[width=80mm]{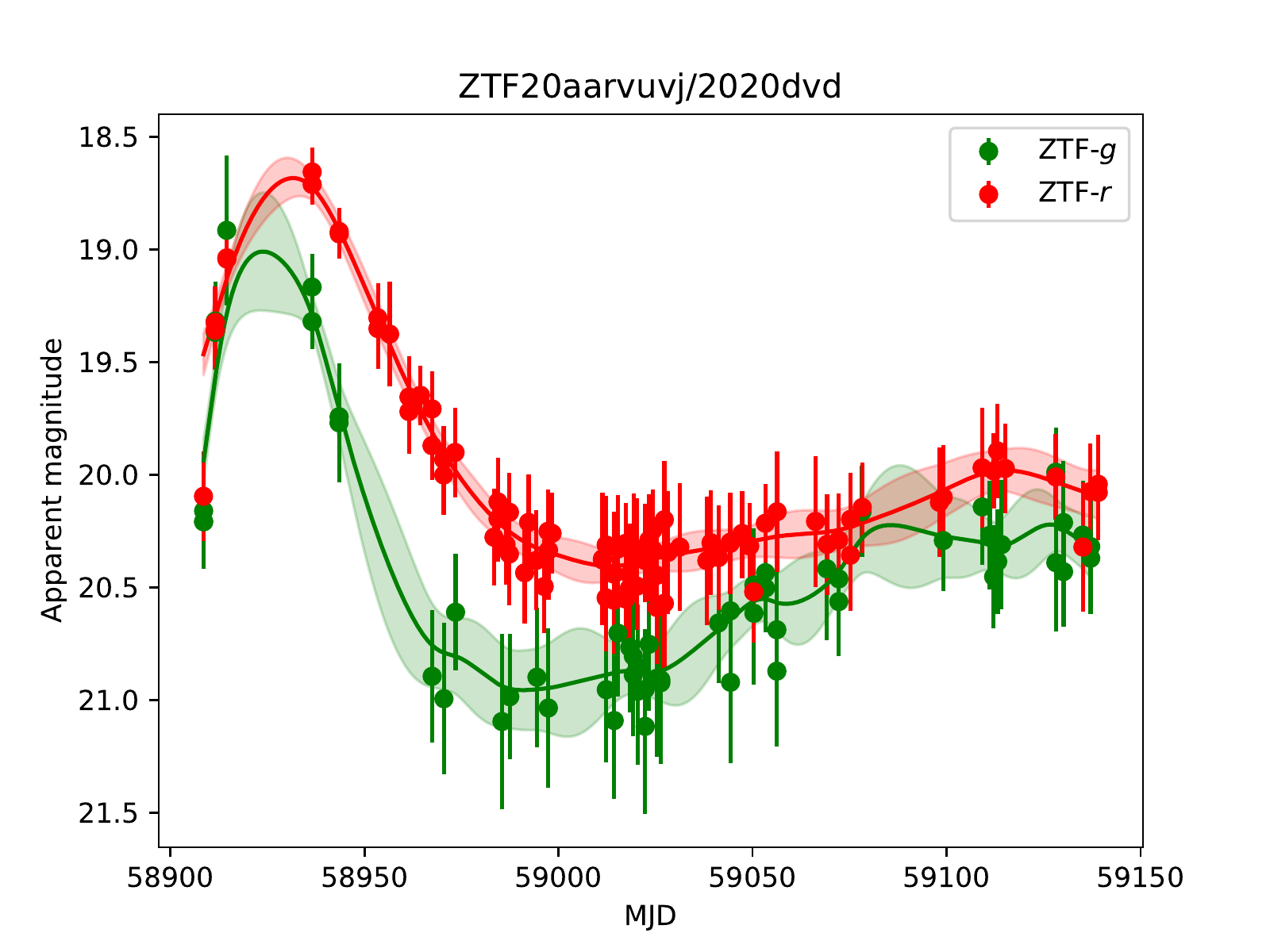}
    \includegraphics[width=80mm]{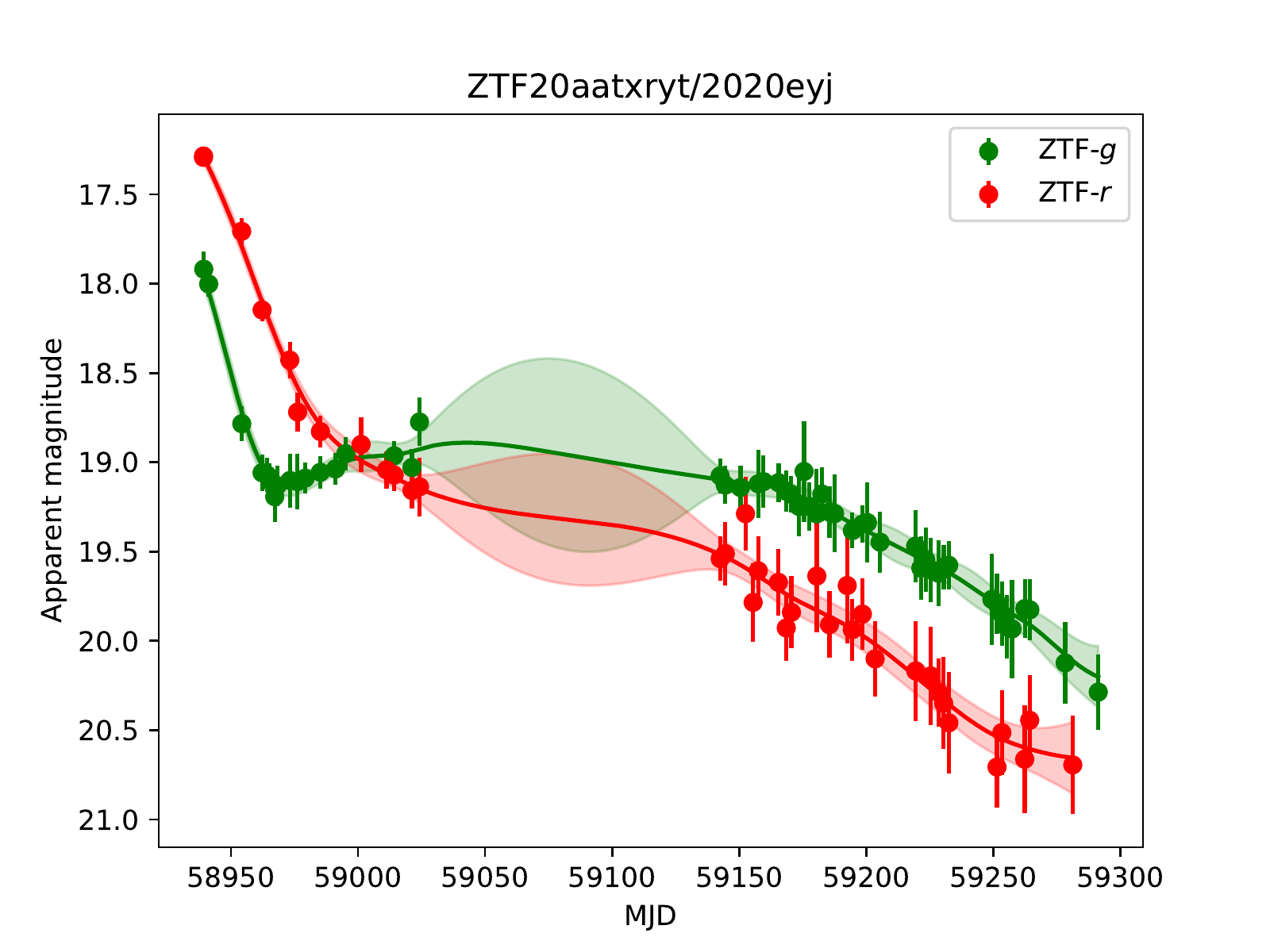}\hfill\includegraphics[width=80mm]{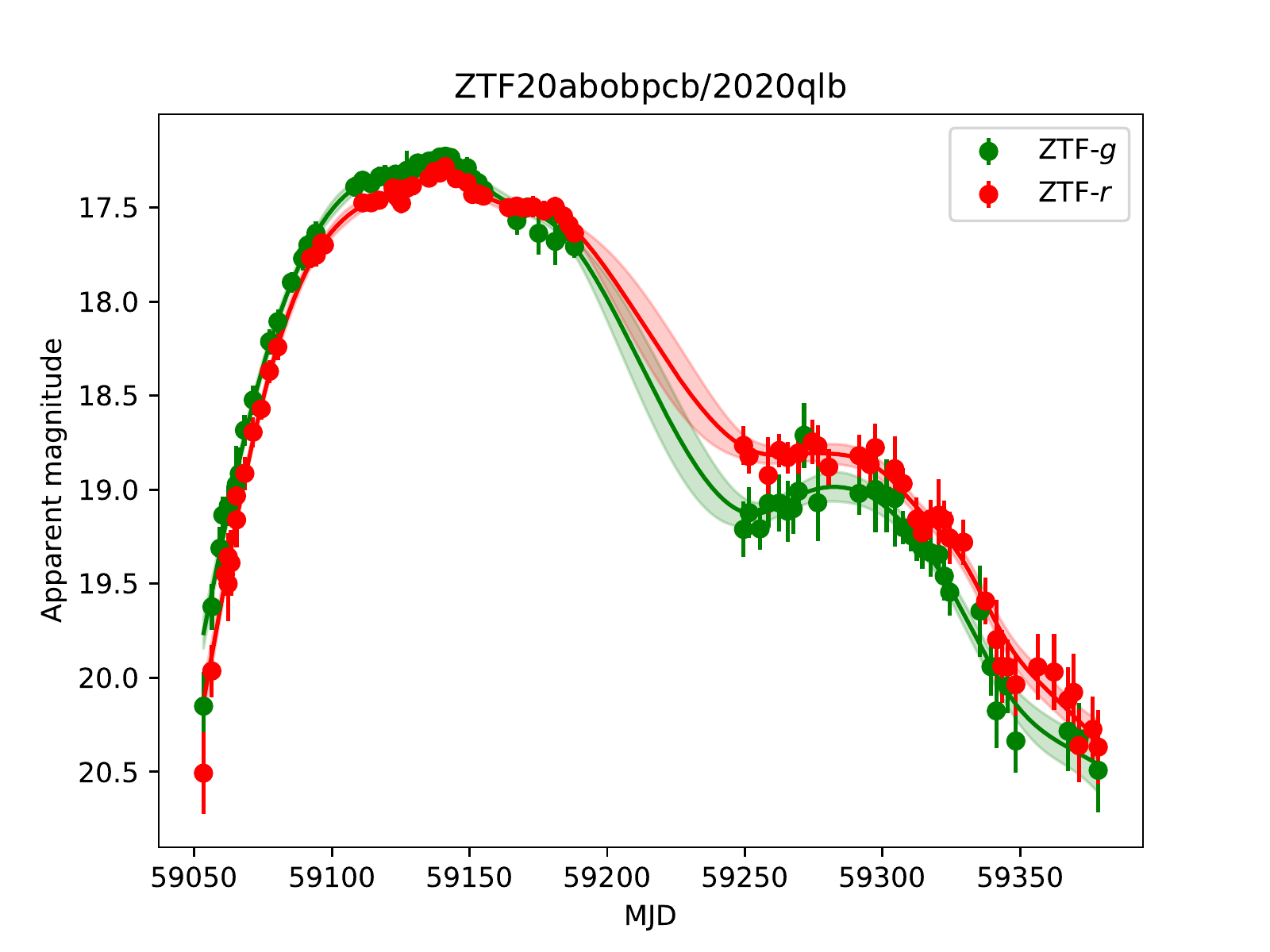}
    \includegraphics[width=80mm]{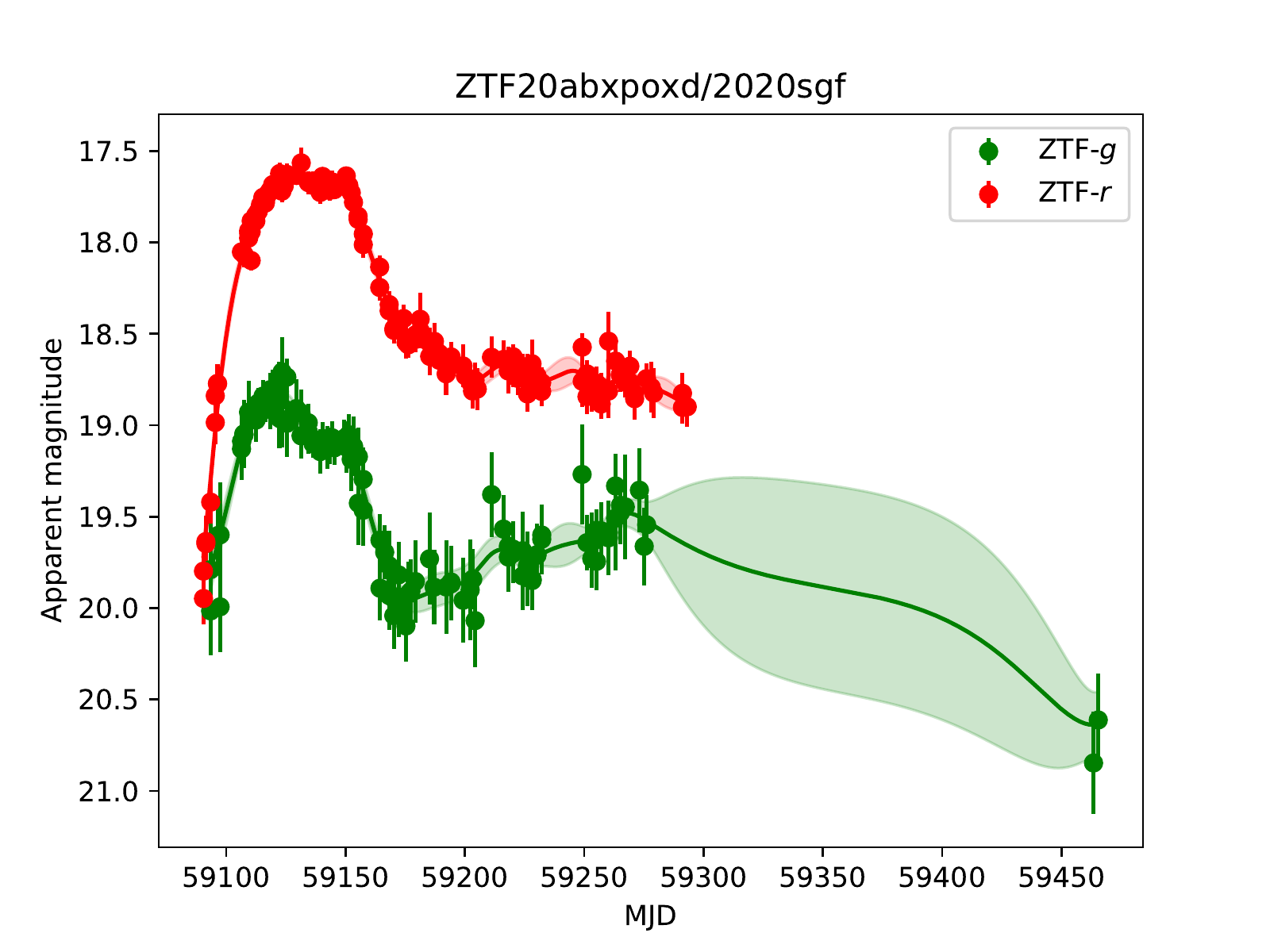}\hfill\includegraphics[width=80mm]{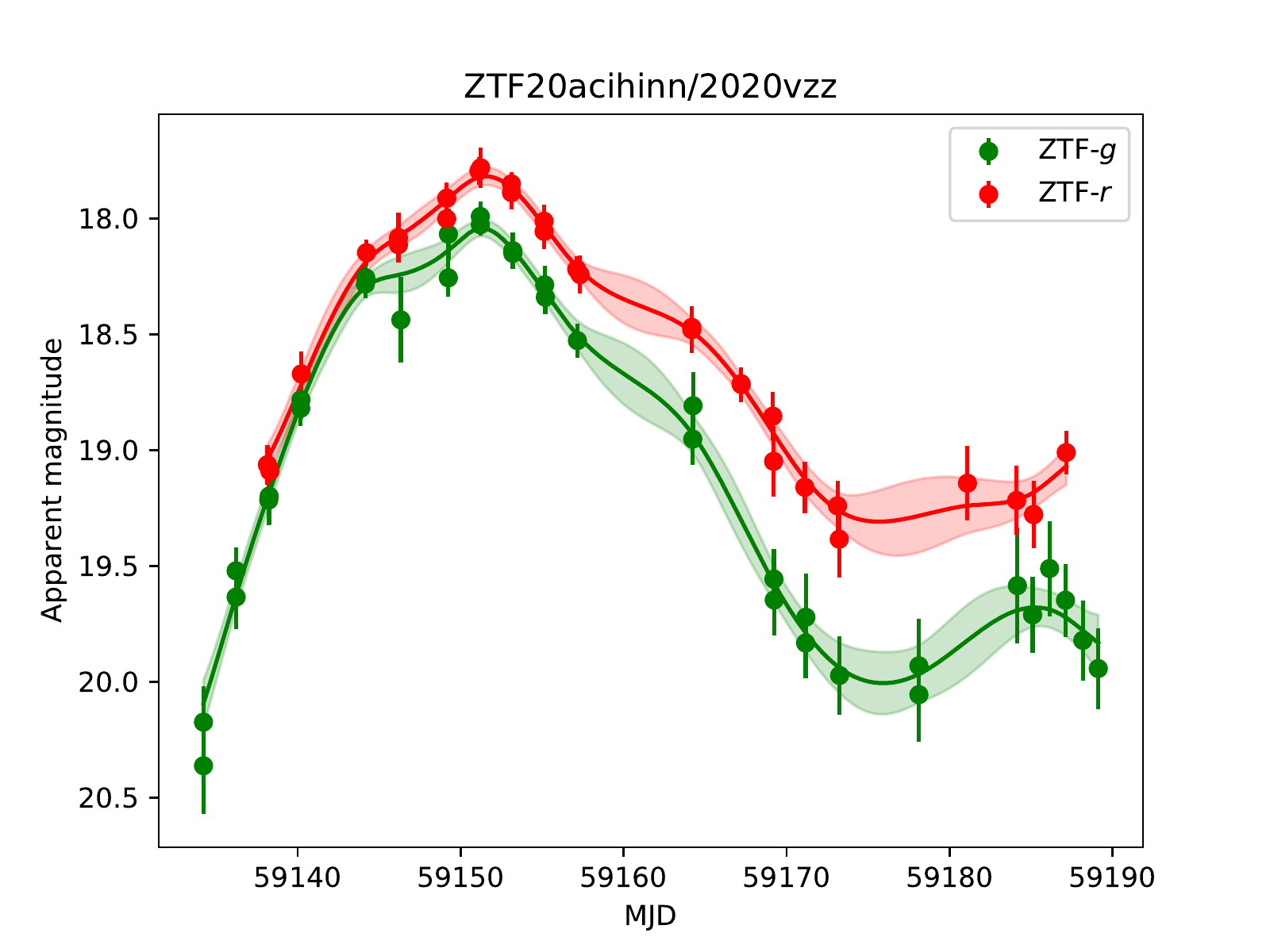}
    \caption{continued}
\end{figure*}

\begin{figure*}
    \ContinuedFloat
    \centering
    \includegraphics[width=80mm]{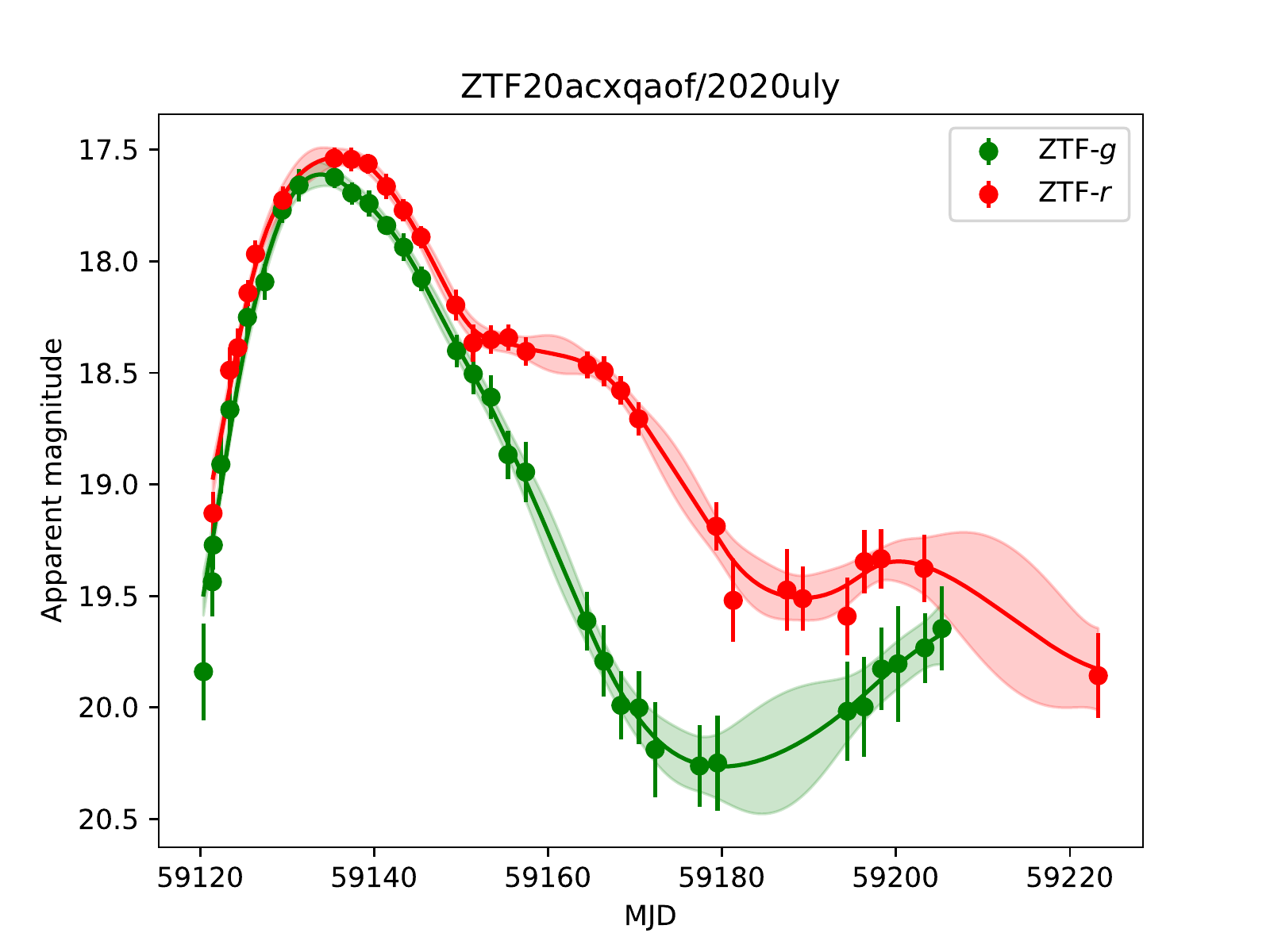}\hfill\includegraphics[width=80mm]{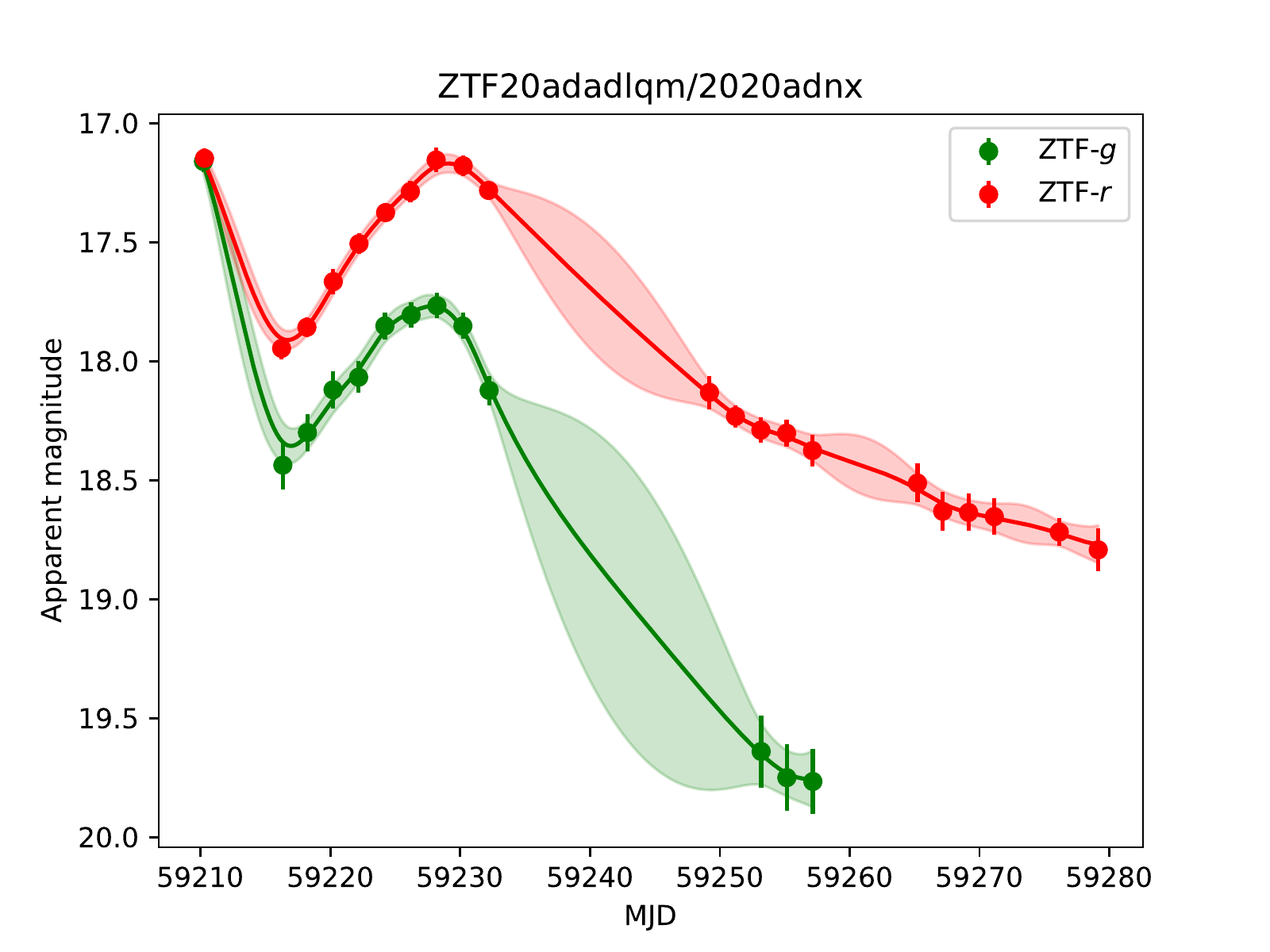}
    \includegraphics[width=80mm]{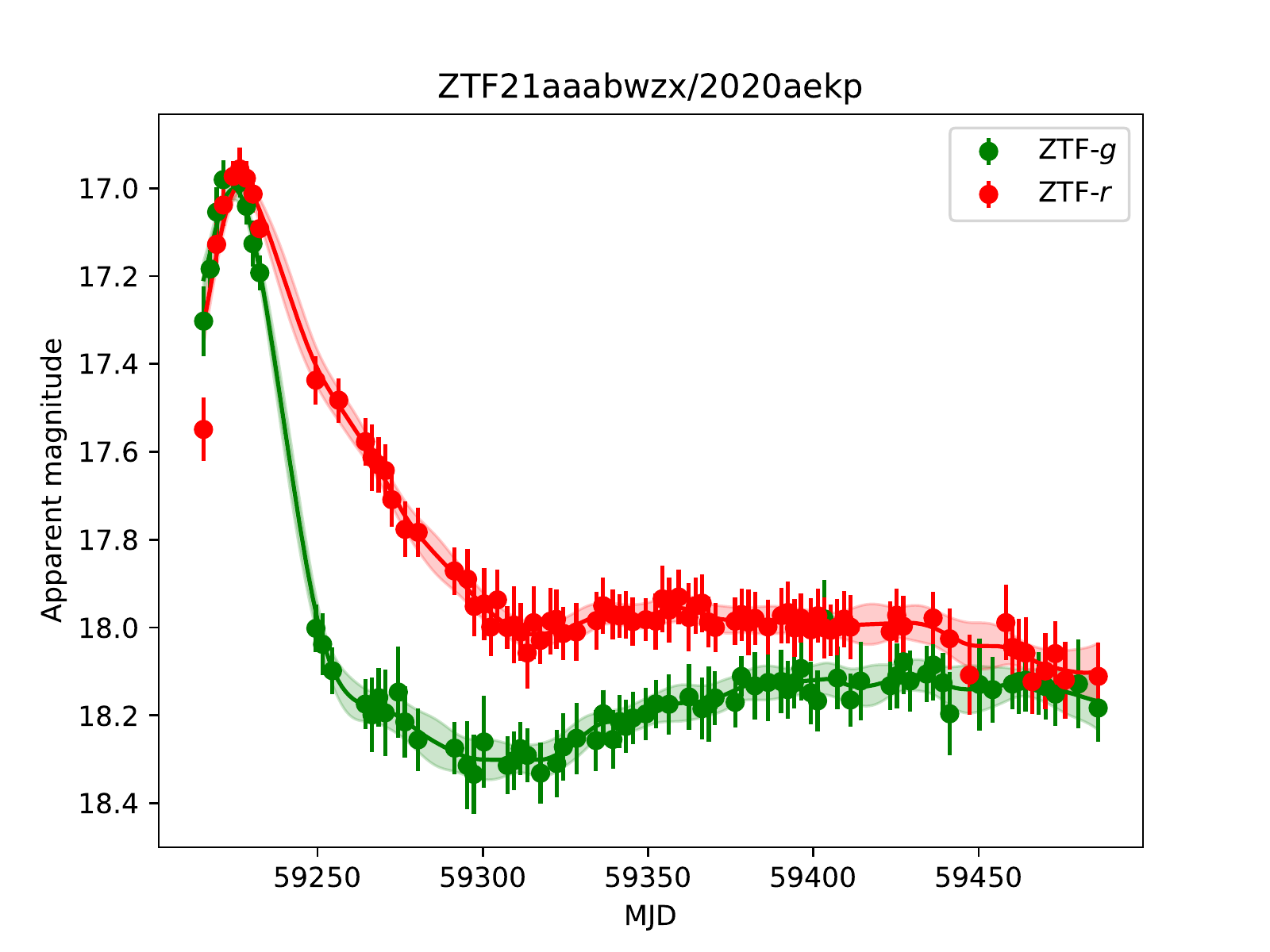}\hfill\includegraphics[width=80mm]{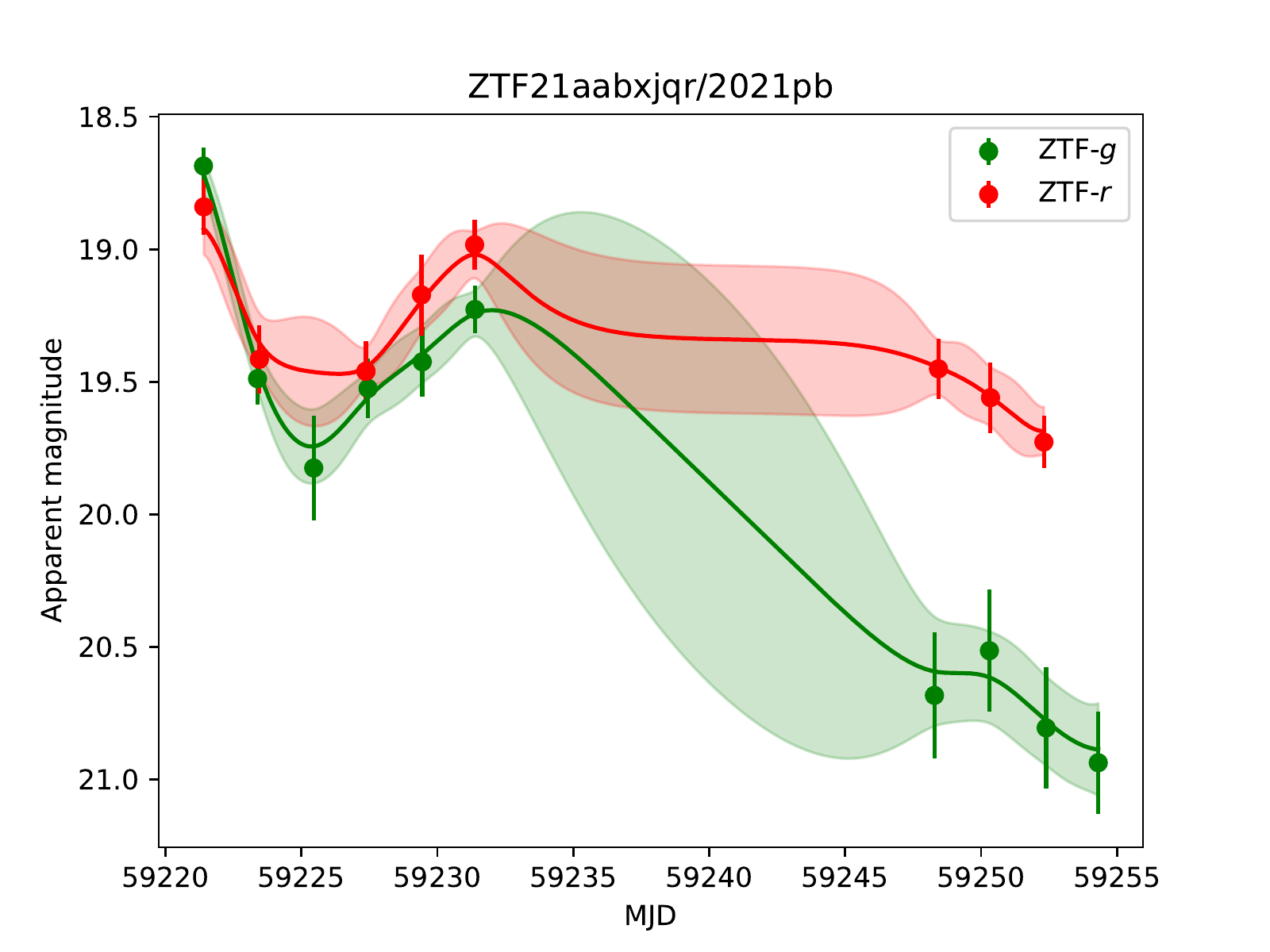}
    \includegraphics[width=80mm]{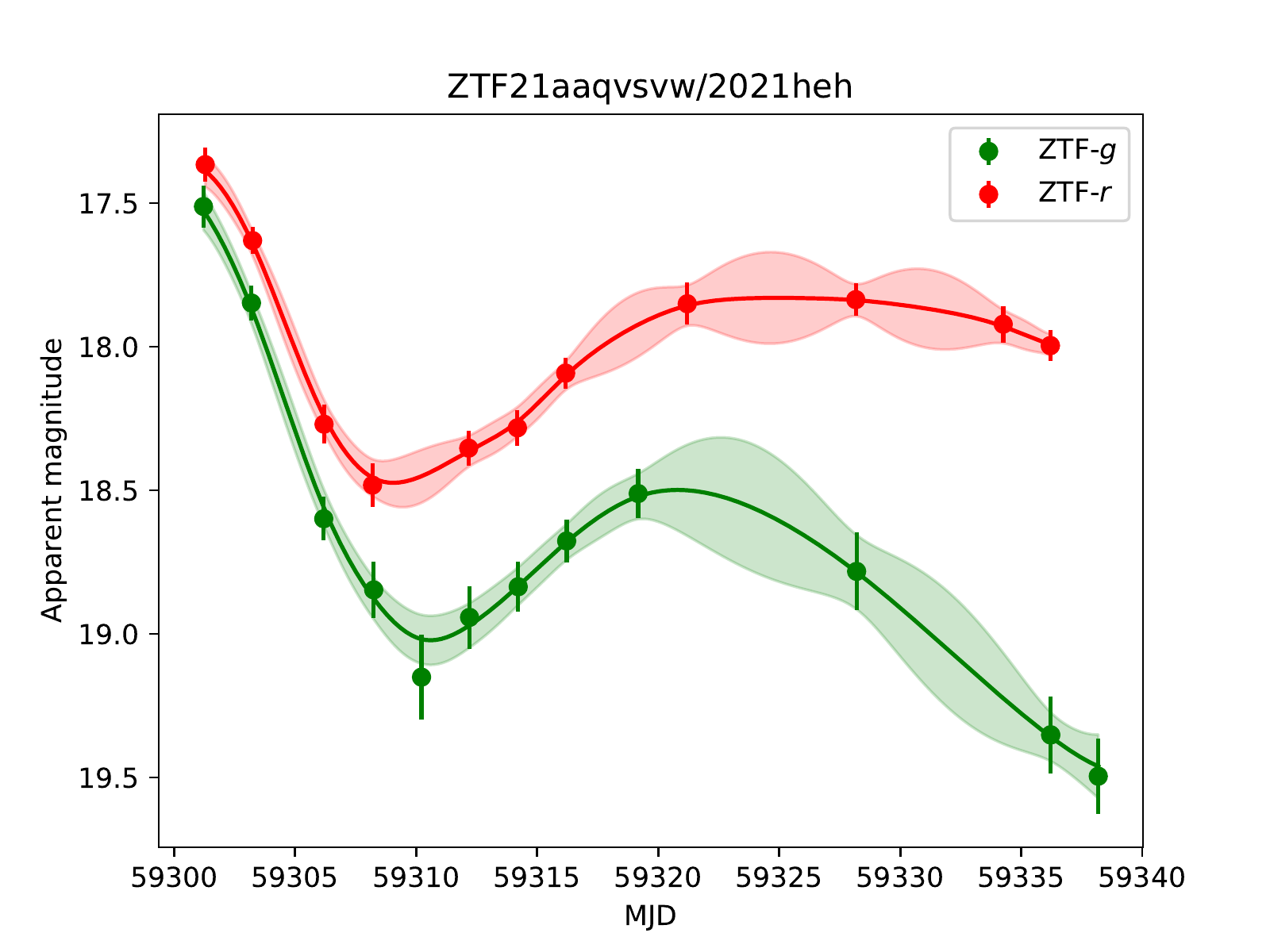}\hfill\includegraphics[width=80mm]{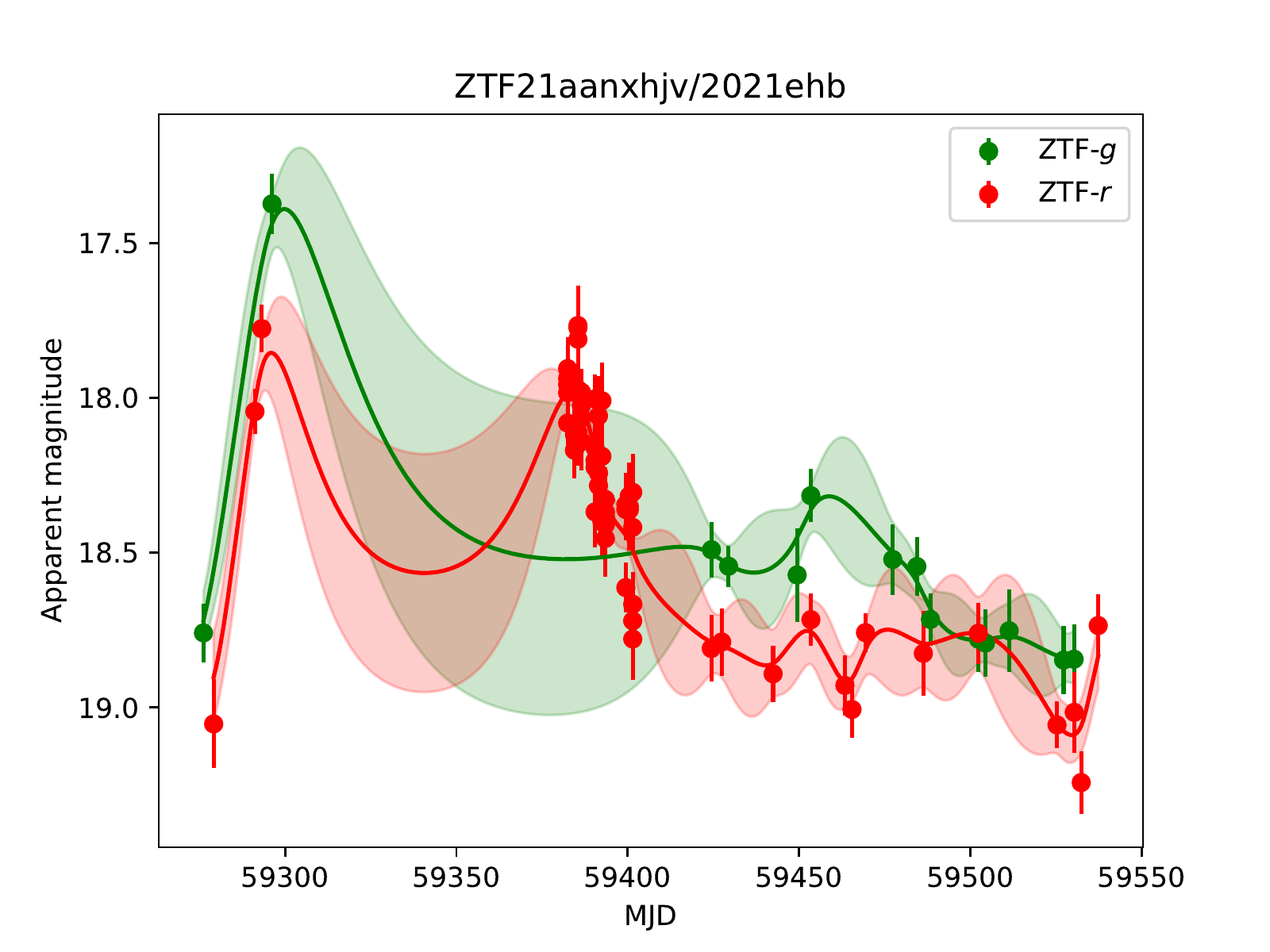}
    \caption{continued}
\end{figure*}

\begin{figure*}
    \ContinuedFloat
    \centering
    \includegraphics[width=80mm]{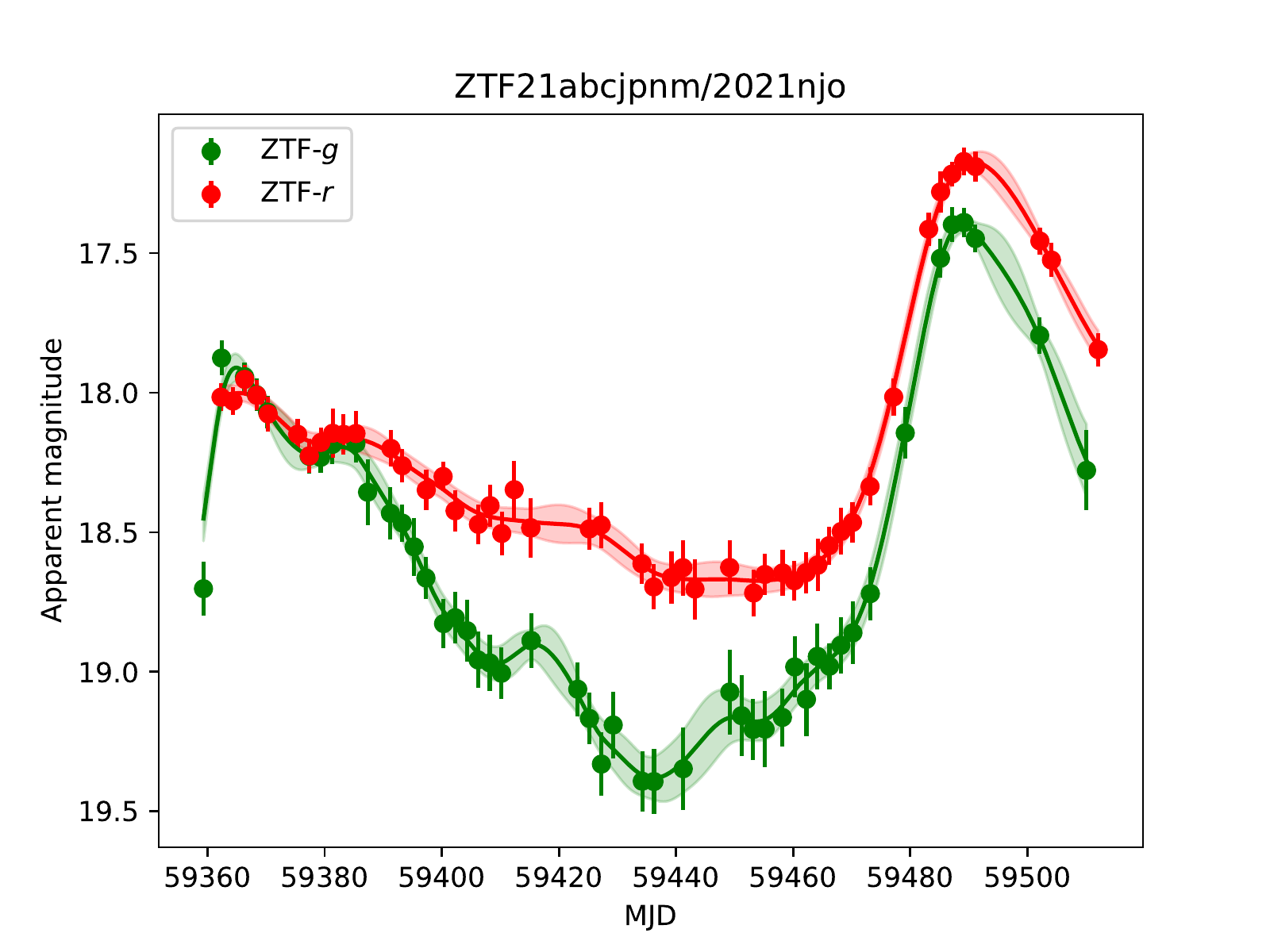}\hfill\includegraphics[width=80mm]{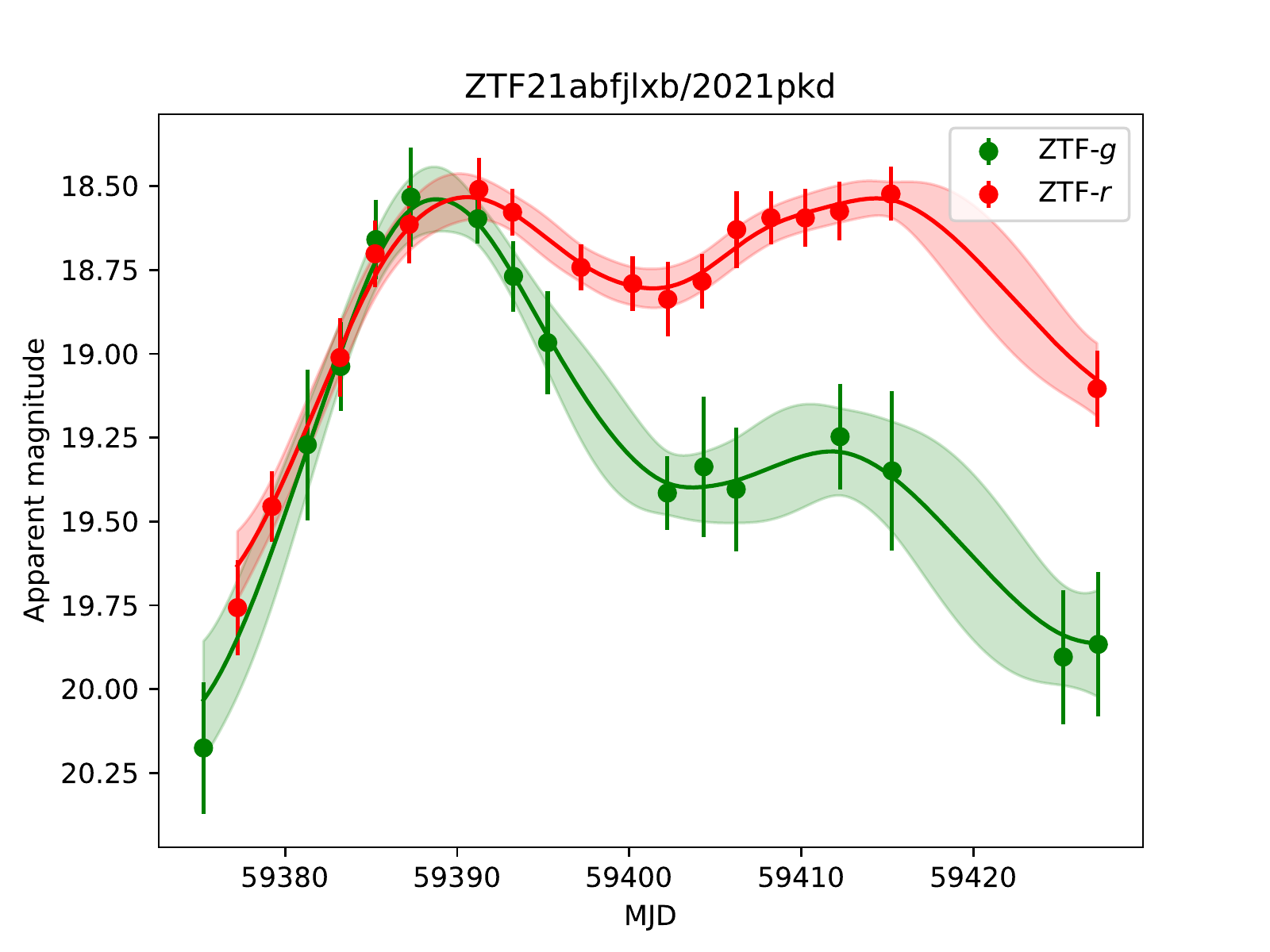}
    \includegraphics[width=80mm]{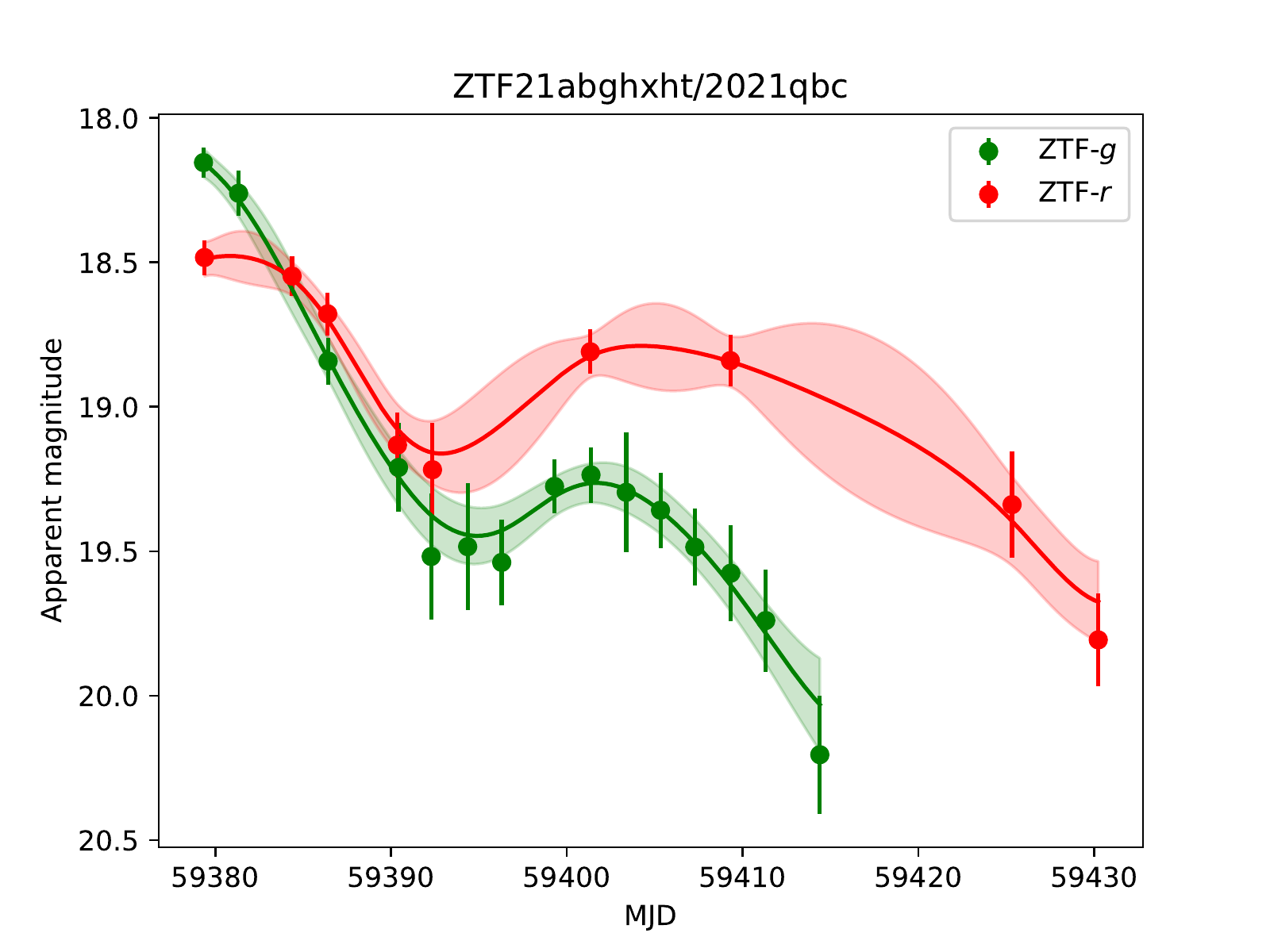}\hfill\includegraphics[width=80mm]{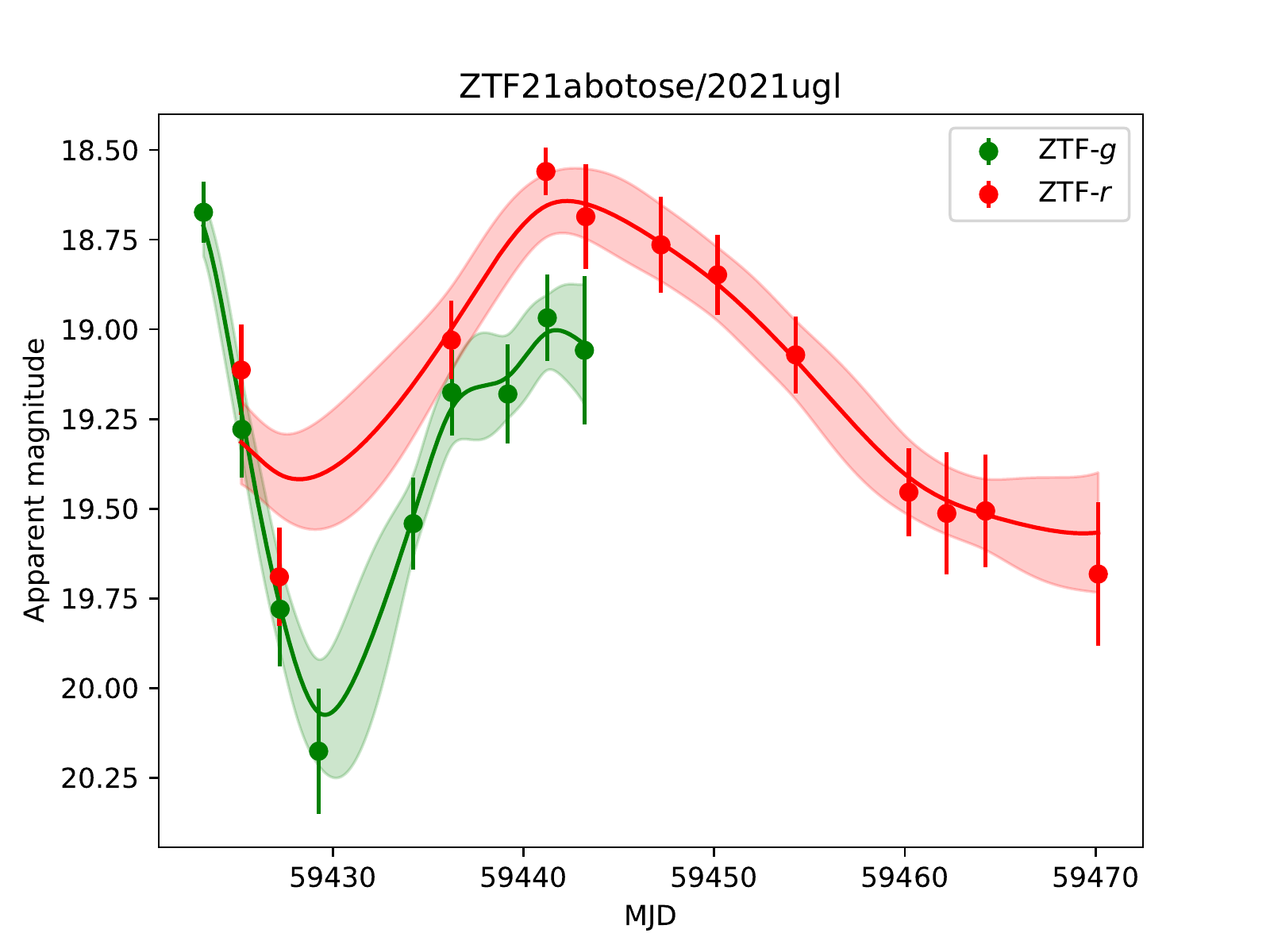}
    \caption{continued}
\end{figure*}

\end{document}